\documentclass[10pt,journal]{IEEEtran}
\usepackage{algpseudocode}                                 % new algorithm package
\usepackage{algorithm}
\usepackage{graphicx}                                      % include pdf figures
\usepackage{amsmath}
\usepackage{amssymb}
\usepackage{amsfonts}
\usepackage{booktabs}
\usepackage{multirow}
\usepackage[subrefformat=parens,farskip=0pt,justification=centering]{subfig}
\usepackage{color}
\usepackage{cite}                                          % more citations in one bracket
\usepackage{comment}                                       % use comment
\usepackage{soul}                                          % use highlight command \hl{}
\soulregister\cite7
\soulregister\ref7
\soulregister\pageref7
\usepackage{amsthm}
\usepackage{etoolbox}                                      % commands \newtoggle, \toggletrue, \iftoggle
\usepackage{url}
\usepackage{nth}                                           % nth command
\usepackage{bm}                                            % bm command
\usepackage{courier}
\usepackage{balance}
\usepackage{threeparttable}
\usepackage[bookmarks=false]{hyperref}
\hypersetup{
    colorlinks = true,
    citecolor  = blue,
    linkcolor  = blue,
    urlcolor   = blue,
}
\usepackage{tikz}
\usetikzlibrary{positioning, fit}
\usepackage{filecontents}                                  % support to pgfplots
\usepackage{pgfplots}
\usepackage{pgfplotstable}
\pgfplotsset{compat=newest}
\usepackage{caption}
\usepackage{cleveref}
\Crefformat{figure}{Fig.~#2#1#3}                           % "Fig.", instead of "Figure"
\Crefname{subfigure}{Fig.}{Figs.}
\Crefformat{table}{TABLE~#2#1#3}                           % "TABLE", instead of "Table"
\definecolor{CUHKorange}{RGB}{244,106,18} %F47012
\definecolor{CUHKblue}{RGB}{0,111,190}    %006FBE
\definecolor{CUHKgreen}{RGB}{0,127,128}   %007F80
\definecolor{CUHKred}{RGB}{228,46,36}     %E42E24
\definecolor{CUHKyellow}{RGB}{198,148,34} %C69422
\definecolor{CUHKdark}{RGB}{114,44,114}   %722C72
\definecolor{CUHKmiddle}{RGB}{144,44,144} %902C90
\definecolor{CUHKlight}{RGB}{167,44,167}

\newtheorem{mydefinition}{\textbf{Definition}}

\algrenewcommand\textproc{\texttt}

\makeatletter
\let\OldStatex\Statex
\renewcommand{\Statex}[1][3]{%
  \setlength\@tempdima{\algorithmicindent}%
  \OldStatex\hskip\dimexpr#1\@tempdima\relax
}
\makeatother

\RequirePackage[normalem]{ulem} %DIF PREAMBLE
\RequirePackage{color}\definecolor{RED}{rgb}{1,0,0}\definecolor{BLUE}{rgb}{0,0,1} %DIF PREAMBLE
\usepackage{psfrag}
\usepackage{xcolor}
\usepackage{listings}
\graphicspath{{./figfiles/}}
\usepackage[outdir=./figfiles/]{epstopdf}
\usepackage{xcolor,colortbl}

\IEEEoverridecommandlockouts
\newcommand{\calO}{\mathcal{O}}

\definecolor{myorange}{RGB}{238,97,42}     %EE612A
\definecolor{mypink}{RGB}{255,171,164}     %FFABA4
\definecolor{mygray}{RGB}{240,240,240}     %F0F0F0
\definecolor{myblue}{RGB}{66,101,175}      %4265AF
\definecolor{lightblue}{RGB}{221,239,250}  %DDEFFA
\definecolor{myred}{RGB}{228,46,36}        %E42E24
\definecolor{mygreen}{RGB}{142,202,206}    %8ECACE
\definecolor{myyellow}{RGB}{255,253,181}   %FFFDB5

\begin{document}

\title{
Cross-layer Optimization for High Speed Adders:\\ A Pareto Driven Machine Learning Approach
}

\author
{
Yuzhe Ma,
Subhendu Roy,
Jin Miao,
Jiamin Chen,
and
Bei Yu
\thanks{
The preliminary version has been presented at the IEEE International Symposium on Low Power Electronics and Design (ISLPED) in 2017.
This work is supported in part by The Research Grants Council of Hong Kong SAR (Project No.~CUHK24209017) and CUHK Undergraduate Summer Research Internship 2017.}
\thanks{Y.~Ma, J.~Chen and B.~Yu are with the Department of Computer Science and Engineering, The Chinese University of Hong Kong, NT, Hong Kong.}
\thanks{S.~Roy is with Intel Corporation, San Jose, CA, USA.}
\thanks{J.~Miao is with the Cadence Design Systems, San Jose, CA, USA.}
}

\maketitle
\thispagestyle{empty}

\iftrue
\begin{abstract}
In spite of maturity to the modern electronic design automation (EDA) tools, 
optimized designs at architectural stage may become sub-optimal after going through physical design flow. 
%Adder design has been such a long studied fundamental problem in VLSI industry. 
Adder design has been such a long studied fundamental problem in VLSI industry yet designers cannot achieve optimal solutions by running EDA tools on the set of available prefix adder architectures. 
In this paper, we enhance a state-of-the-art prefix adder synthesis algorithm to obtain a much wider solution space in architectural domain.
On top of that, a machine learning-based design space exploration methodology is applied to predict the Pareto frontier of the adders in physical domain,
which is infeasible by exhaustively running EDA tools for innumerable architectural solutions.
Considering the high cost of obtaining the true values for learning,
an active learning algorithm is proposed to select the representative data during learning process, which uses less labeled data  while achieving better quality of Pareto frontier.
Experimental results demonstrate that our framework can achieve Pareto frontier of high quality over a wide design space,
bridging the gap between architectural and physical designs.
Source code and data are available at \url{https://github.com/yuzhe630/adder-DSE}.
\end{abstract}
\fi

\section{Introduction} \label{sec:intro}
%% General Paragraph  -- more high-level
\IEEEPARstart{I}{n}
the last decades, the industrial EDA tools have advanced towards optimality, especially at the individual stages of VLSI design cycle.
Nevertheless, with growing design complexity and aggressive technology scaling, physical design issues have become more and more complex.
As a result, the constraints and the objectives of higher layers, such as the system or logic level, are very difficult to be mapped into those of lower layers, such as physical design, and vice-versa,
thereby creating a \textit{gap} between the optimality at the logic stage and the physical design stage.
This necessitates the innovation of data-driven methodologies, such as machine learning \cite{PD_ASPDAC2015_Yu,LEARN_ASPDAC2016_Chan,LEARN_DATE2016_Meng,PDN_ISPD2016_Chang,CTS_ISPD2008_Samanta}, to bridge this gap.

\begin{figure}[tb]
    \centering
    \includegraphics[width=0.46\textwidth]{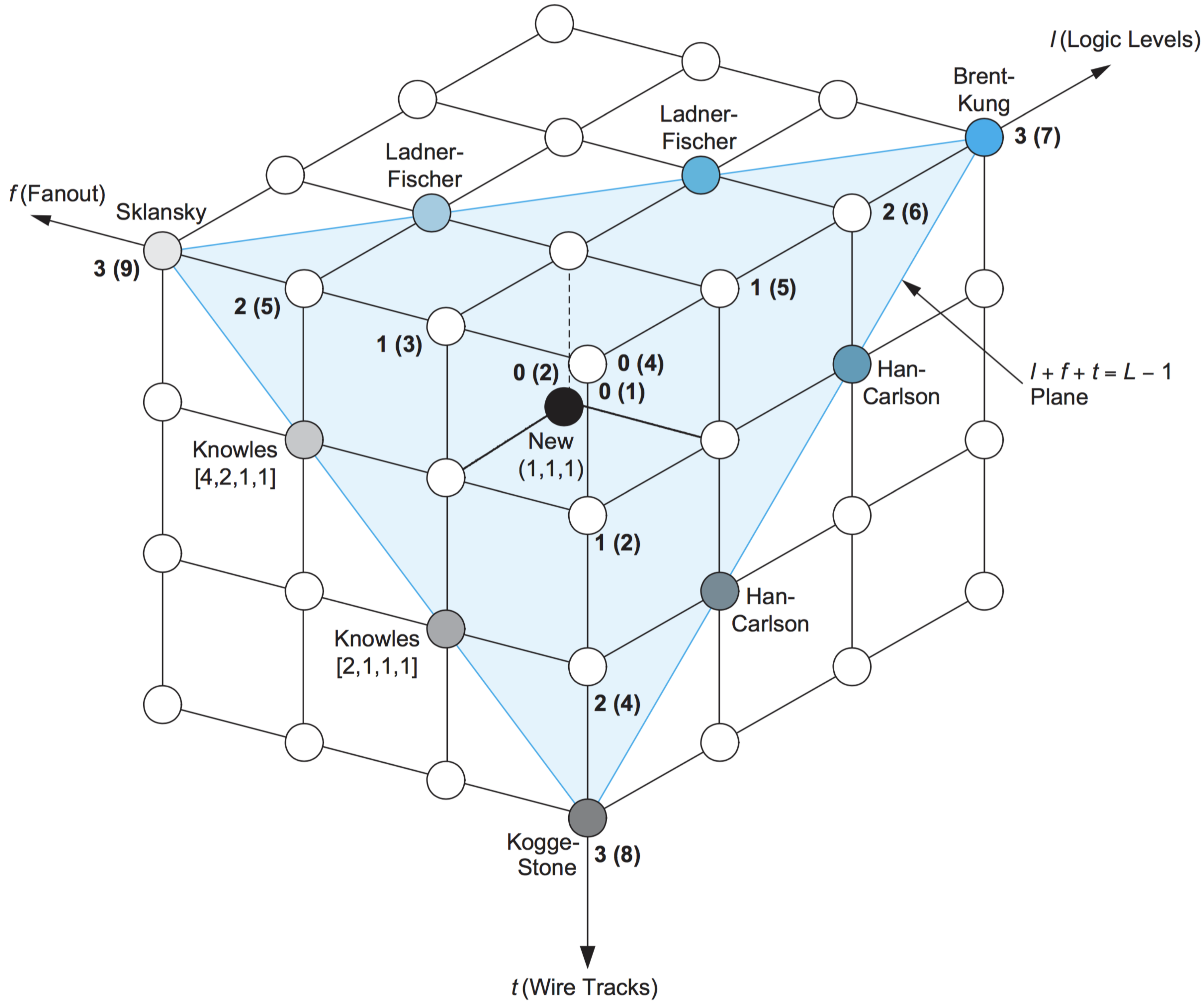}
    \caption{Regular adders (picture taken from \cite{VLSI_B2004_Weste}).}
    \label{fig:regular}
\end{figure}

%% Diving into adder
Adder design is one of the fundamental problems in digital semiconductor industry, 
and its main bottleneck (in terms of both delay and area) is the carry-propagation unit. 
This unit can be realized by hundreds of thousands of parallel prefix structures, but it is hard to evaluate the final metrics without running through physical design tools.
Historically, regular adders~\cite{ADDER_TOC1982_BrentKung,ADDER_TOC1973_KoggeStone,ADDER_ARITH1987_HanCarlson,ADDER_IRE1960_Sklansky}
have been proposed for achieving the corner points in terms of various metrics as shown in \Cref{fig:regular} in architectural stage.
The main motivation for structural regularity was the ease of manual layout, but EDA tools now taking care of all physical design aspects, the regularity is no longer essential.
Moreover, the extreme corners do not map well to the physical design metrics after synthesis, placement and routing.
To address this gap between prefix adder synthesis and actual physical design of the adders, custom adders are typically designed by tuning parameters,
such as gate-sizing, buffering etc., targeting at the optimization of power/performance metrics for a specific technology library
\cite{ADDER_CICC2009_Zhou,ADDER_ASPDAC2007_Liu}.
However, this custom approach (i) needs significant engineering effort, (ii) is not flexible to Engineering Change Order (ECO),
and (iii) does not guarantee the optimality.
	   
The algorithmic synthesis approach resolves the first two issues of the custom approach, by adding more flexibility to the late ECO changes and reducing the engineering effort. 
Based on the number of solutions, the existing adder synthesis algorithms can be broadly classified into two categories. 
The first and the most common approach is to generate a single prefix network for a set of structural constraints, such as the logic level, fan-out etc. 
Several algorithms have been proposed to minimize the size of the prefix graph ($s$) under given bit-width ($n$) and logic-level ($L$) constraints
\cite{ADDER_GLSVLSI2007_Matsunaga,ADDER_ICCAD2003_Liu,ADDER_DAC1990_Fishburn,ADDER_IWLAS1996_Zimmermann}.
Closed form theoretical bounds for size-optimality are provided by \cite{ADDER_JALG1986_Snir} for $L \geq 2 \log_2 n - 2$.
\cite{ADDER_ASPDAC2005_Zhu} has given more general bound for prefix graph size, but when $L$ is reduced to $\log_2 n$, a pre-requisite for high-performance adders, there is no closed form bound for $s$. 
\cite{ADDER_ASPDAC2015_Roy} presents a polynomial-time algorithm for generating prefix graph structures by restricting both logic-level and fan-out. 
The limitations in these approaches are two-fold, (i) this restricted set of structures is not capable of exploring the large solution space, 
and (ii) since it is very hard to analytically model the physical design complexities, such as wire-length and congestion issues, the physical design metrics, 
such as the area, power, delay etc., may not be mapped well to the prefix structure metrics, such as the size, max-fan-out ($mfo$) etc. 
This motivates the second category of algorithms where thousands of prefix adder solutions can be generated and explored for synthesis and physical design in the commercial EDA tools.

One such approach is \cite{ADDER_TCAD2014_Roy} which presents an exhaustive bottom-up enumeration technique with several pruning strategies to generate innumerable prefix structure solutions. 
However, it has two issues, (i) this approach cannot provide solutions in several cases for restricted fan-out, which can control the congestion and load-distribution during physical design \cite{ADDER_ASPDAC2015_Roy}. 
As a result, it may still miss the \textit{good} solution space to a large extent, and (ii) it is computationally very intensive to run all solutions through synthesis, placement and routing. 

In this paper, we enhance the algorithm in~\cite{ADDER_TCAD2014_Roy,ADDER_USP2014_Choudhury} to generate adders under any arbitrary $mfo$ constraint, which enables a \textit{wider} adder solution space in \textit{logical} form. 
To tackle the high computational effort during the physical design flow, we further propose to use machine learning to perform the design space exploration in \textit{physical} solution space.
We develop Pareto frontier driven machine learning methodologies to achieve rich adder solutions with trade-offs among power, area, and delay.
As a passive supervised learning, the proposed \textit{quasi-random} sampling approach is able to select representative prefix adders out of the hundreds of thousands of prefix structures.

It should be noted that various machine learning algorithms have been investigated to explore design space in different design scenarios.
Palermo \textit{et al}.~\cite{CAD_TCAD2009_Palermo} deploy both linear regression and artificial neural network for multiprocessor systems-on-chip design.
Lin \textit{et al}.~\cite{CAD_DAC2013_Liu} present a random forest-based learning model in high level synthesis, which can find an approximate Pareto-optimal designs effectively.
Meng \textit{et al}.~\cite{CAD_DATE2016_Meng} propose a random forest-based method for Pareto frontier exploration,
where non-Pareto-optimal designs are carefully eliminated through an adaptive strategy.
Multiple predictions can be obtained through random forest, which can be used for estimating the uncertainty.
Superior to the random forest, in this paper we further propose an active learning approach based on Gaussian Process (GP), which by nature can estimate the prediction uncertainty efficiently.

%To the best of our knowledge, this is the first attempt to bridge architectural synthesis with physical design in the field of EDA using machine learning.
Our main contributions are summarized as follows:
\begin{itemize}
  \item
  A comprehensive framework for optimal adder search by machine learning methodology bridging the prefix architecture synthesis to the final physical design;
  \item
  An enhancement to a state-of-the-art prefix adder algorithm~\cite{ADDER_TCAD2014_Roy} to optimize the prefix graph size for restricted fan-out and explore a wider solution space;
  \item
  A machine learning model for prefix adders, guided by quasi-random data sampling with features considering architectural attributes and EDA tool settings;
  \item
  A design space exploration method to generate the Pareto frontier for delay vs.~power/area over a wide design space;
  \item
  An active learning approach for the design space exploration, which uses less labeled data and achieves better quality of Pareto frontier.
\end{itemize}   

The rest of the paper is organized as follows. 
%First we give a brief background of the parallel prefix adders in Section \ref{sec:prelim}. 
%Section \ref{sec:pgstr} presents the techniques for enhancing a state-of-the-art algorithm \cite{ADDER_TCAD2014_Roy} to scale for restricted fan-out leading to wider solution space. 
\Cref{sec:prefixadder} presents the background of prefix adder synthesis,
while \Cref{sec:mladder} discusses our prefix graph generation algorithm.
Next, two machine learning approaches of design space exploration for high-performance adders are described.
\Cref{sec:psl} presents the passive supervised learning, while \Cref{sec:pal} introduces a Pareto frontier driven active learning approach.
\Cref{sec:result} lists the experimental results, followed by conclusion in \Cref{sec:conclu}.

\newcommand{\twopartdef}[4]
{
    \left\{
        \begin{array}{ll}
            #1 & \mbox{if } #2 \\
            #3 & \mbox{} #4
        \end{array}
    \right.
}

\section{Prefix Adder Synthesis}
\label{sec:prefixadder}

In this section, we first provide the background of the prefix adder synthesis problem.
Then we present a brief discussion on the algorithm presented in \cite{ADDER_TCAD2014_Roy},
which we enhance to our \textbf{Prefix Graph Generation (PGG)} algorithm to synthesize the prefix adder network.

\subsection{Preliminaries}
\label{sec:prelim}
%{{{
An $n$ bit adder accepts two $n$ bit addends $A = a_{n-1}..a_{1}a_{0}$ and $B = b_{n-1}..b_{1}b_{0}$ as input, and computes the output sum $S = s_{n-1}..s_{1}s_{0}$ and carry out $C_{out} = c_{n-1}$, 
where $s_{i} = a_{i} \oplus b_{i} \oplus c_{i-1}$ and $c_{i} = a_{i}b_{i} + a_{i}c_{i-1} + b_{i}c_{i-1}$. 
The simplest realization for the adder network is the ripple-carry-adder, 
but with logic level $n-1$, which is too slow. For faster implementation, carry-lookahead principle is used to compute the carry bits. 
Mathematically, this can be represented with bitwise (group) generate function $g$ ($G$) and propagate function $p$ ($P$) by the Weinberger's recurrence equations as follows \cite{ADDER_ARITH2005_Zeydel}:
\begin{itemize}
  \item Pre-processing (inputs): Bitwise generation of $g$, $p$
\begin{eqnarray}
g_{i} = a_{i} \cdot b_{i}~\text{and}~p_{i} = a_{i} \oplus b_{i}.
\end{eqnarray}

  \item Prefix processing: This part is the main carry-propagation component where the concept of generate/propagate is extended to multiple bits and $G_{[i:j]}$, $P_{[i:j]}$ ($i \geq j$) are defined as
\begin{align}
  P_{[i:j]} & = \twopartdef {p_{i},} {i=j,} {P_{[i:k]} \cdot P_{[k-1:j]},}             {\textrm{otherwise},}\\ %\nonumber
  G_{[i:j]} & = \twopartdef {g_{i},} {i=j,} {G_{[i:k]} + P_{[i:k]} \cdot G_{[k-1:j]},} {\textrm{otherwise}.}
\end{align}
   The associative operation $\circ$ is defined for ($G$, $P$) as:
\begin{align}
\begin{split}
(G,P)_{[i:j]} &= (G,P)_{[i:k]} \circ (G,P)_{[k-1:j]} \\
              &= (G_{[i:k]} + P_{[i:k]} \cdot G_{[k-1:j]}, P_{[i:k]} \cdot P_{[k-1:j]}).
\end{split}
\end{align}
   \item Post-processing (outputs): Sum/Carry-out generation
\begin{eqnarray}
s_{i} = p_{i} \oplus c_{i-1}, \quad c_{i} = G_{[i:0]}, ~\text{and}~C_{out} = c_{n-1}.
\end{eqnarray}
\end{itemize}

\begin{figure}[tb!]
    \centering
    \begin{minipage}{.42\linewidth}
        \includegraphics[height=2.8cm]{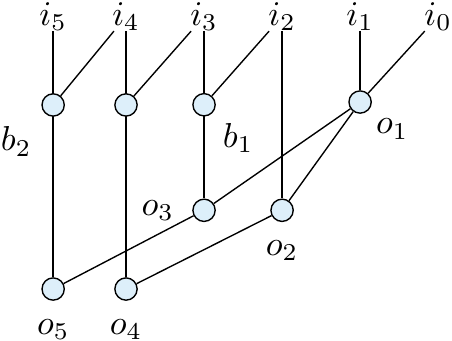}
        \caption{$6$ bit prefix adder network.}  \label{fig:prelim}
    \end{minipage}
    \begin{minipage}{.56\linewidth}
        \includegraphics[height=2.2cm]{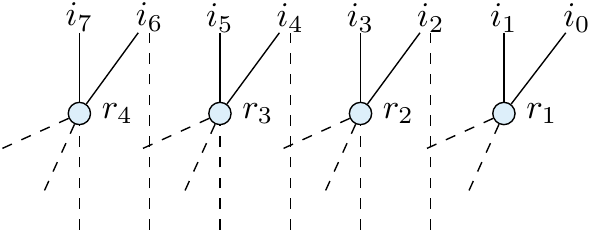}
        \caption{Imposing semi-regularity.}
        \label{fig:semiregular}
    \end{minipage}
\end{figure}

The `Prefix processing' or carry propagation network can be mapped to a prefix graph problem with inputs $i_{k}=(p_{k},g_{k})$ and outputs $o_{k} = c_{k}$, 
such that $o_{k}$ depends on all previous inputs $i_{j}$ ($j \leq k$). Any node except the input nodes is called a \textit{prefix} node. Size of the prefix graph is defined as the number of prefix nodes in the graph. 
\Cref{fig:prelim} shows an example of such prefix graph of $6$ bit and we can see that $C_{out}$ = $c_{5}$ = $o_{5}$ is given by 
\begin{equation}
o_{5}  = (i_{5} \circ i_{4}) \circ ((i_{3} \circ i_{2}) \circ (i_{1} \circ i_{0})). \label{asso}
%\\ \nonumber
%       &=& \ o \ ((x_{3} \ o \ x_{2}) \ o \ (x_{1} \ o \ x_{0})) \label{asso}
\end{equation}

\iffalse
%{{{
\begin{figure}[tb!]
%\vspace{-10pt}
\begin{center}
\psfrag{x0}[][]{\small $i_{0}$}
\psfrag{x1}[][]{\small $i_{1}$}
\psfrag{x2}[][]{\small $i_{2}$}
\psfrag{x3}[][]{\small $i_{3}$}
\psfrag{x4}[][]{\small $i_{4}$}
\psfrag{x5}[][]{\small $i_{5}$}
\psfrag{x6}[][]{\small $x_{6}$}
\psfrag{x7}[][]{\small $x_{7}$}
\psfrag{o0}[][]{\small $o_{0}$}
\psfrag{o1}[][]{\small $o_{1}$}
\psfrag{o2}[][]{\small $o_{2}$}
\psfrag{o3}[][]{\small $o_{3}$}
\psfrag{o4}[][]{\small $o_{4}$}
\psfrag{o5}[][]{\small $o_{5}$}
\psfrag{b1}[][]{\small $b_1$}
\psfrag{b2}[][]{\small $b_2$}
\psfrag{size}[][]{\small $size = 13$}
\psfrag{level}[][]{\small $level = 3$}
\psfrag{fanout}[][]{\small $mfo = 3$}
\includegraphics[width=0.2\textwidth, height=3cm]{prelim}
%\psfragfig[width=0.2\textwidth, height=3cm]{prelim.eps}
%\vspace{-10pt}

\end{center}
\end{figure}
%}}}
\fi

Size ($s$), logic level ($L$) and maximum-fan-out ($mfo$) for this network are respectively $8$, $3$ and $2$. 
Note that here the number of fan-ins for each of the associative operation $o$ is two, thus this is called radix-$2$ implementation of the prefix graph. 
However, there exist other options such as radix-$3$ or radix-$4$, but the complexity is very high and not beneficial in static CMOS circuits \cite{ADDER_ASILOMAR2011_Ketter}.
In this work, the logic levels for all output bits are $\log_2 n$, \textit{i.e.}, the minimum possible, to target high performance adders. 
%In Equation~\eqref{asso} or Fig.~\ref{prelim}, the number of fan-ins for each of the associative operation $o$ is two and thus it is called as radix-$2$ implementation of prefix network. 
%However, there exist other choices such as radix-$3$ or radix-$4$, but the complexity is very high and not beneficial in static CMOS circuits \cite{KETTER}. 
%In \cite{KAO}\cite{NAFFZIGER} fast domino adders are implemented using radix-$4$ Ling network, but domino logic has been phased out due to the high power consumption. 
%\cite{PATIL} also demonstrates the energy-efficiency of radix-$2$ implementation. 
%Ling adders \cite{ZEYDEL}\cite{NIKOLOS} have been proposed as an alternative in the past by transforming Weinberger's equations to provide better performance, 
%and it is always possible to explore the Ling implementation of any prefix network due to the direct mapping between Weinberger's equations and Ling's equations \cite{ZEYDEL}. 
%}}}

\subsection{Discussion on \cite{ADDER_TCAD2014_Roy}}
\label{sec:pgstr}
%{{{
Our PGG algorithm to generate the prefix graph structures for physical solution space exploration is based on \cite{ADDER_TCAD2014_Roy}.
%which is an exhaustive bottom-up and pruning based enumeration technique for prefix adder synthesis.
%Since our prefix network synthesis is built on top of \cite{ADDER_TCAD2014_Roy}, 
So it is imperative to first discuss about \cite{ADDER_TCAD2014_Roy}. 
However, we omit the details and only mention the key points of \cite{ADDER_TCAD2014_Roy} due to space constraint.

\cite{ADDER_TCAD2014_Roy} is an exhaustive bottom-up and pruning based enumeration technique for prefix adder synthesis.
This work presented an algorithm to generate all possible $n+1$ bit prefix graph structures from any $n$ bit prefix graph.
 Then this algorithm is employed in a bottom-up fashion (from $1$ bit adder to $2$ bit adders, then from all $2$ bit adders to $3$ bit adders, and so on) to synthesize prefix graphs of any bit-width.
 As a result, scalability issue arises due to the exhaustive nature of the algorithm, which is then tackled by adopting various pruning strategies to scale the approach.
%Even with those pruning strategies, \cite{ADDER_TCAD2014_Roy} has been theoretically proved to be size-optimal without any fan-out restriction when the bit-width $n$ is a power of $2$, 
%and logic level is $log_2 n$. 
However, the pruning strategies are not sufficient to scale the algorithm well for different fan-out constraints. 
%This is because with fanout restrictions, the search space is enormous as\cite{ADDER_TCAD2014_Roy} attempts to explore all possible adder solutions of bit-width $n+1$ from the existing adder solutions of bit-width $n$. 
So when it intends to find the solutions for higher bit adders, the intermediate adder solutions that need to be generated are often huge. 
Consequently, it fails to get fan-out restricted (e.g. when $mfo=8,10,12$ etc. for $64$ bit adders) solutions even with $72$GB RAM due to the generation of innumerable intermediate solutions \cite{ADDER_ASPDAC2015_Roy}. 
Pruning strategies, such as size-bucketing \cite{ADDER_TCAD2014_Roy}, help to achieve solutions in some cases, but with sub-optimality. 
So design space-exploration based on this algorithm can miss a significant spectrum of the adder solutions. 
%}}}

\subsection{Our PGG Algorithm}
\label{sec:pgg}

To better explore the wide design space of adders, in this paper we have enhanced \cite{ADDER_TCAD2014_Roy} for different fan-out constraints by incorporating more pruning techniques.

\subsubsection{Semi-Regularity in Prefix Graph Structure}

\iffalse
%{{{
\begin{figure}[tb!]
\begin{center}
\psfrag{x0}[][]{$i_{0}$}
\psfrag{x1}[][]{$i_{1}$}
\psfrag{x2}[][]{$i_{2}$}
\psfrag{x3}[][]{$i_{3}$}
\psfrag{x4}[][]{$i_{4}$}
\psfrag{x5}[][]{$i_{5}$}
\psfrag{x6}[][]{$i_{6}$}
\psfrag{x7}[][]{$i_{7}$}
\psfrag{i1}[][]{$r_{1}$}
\psfrag{i2}[][]{$r_{2}$}
\psfrag{i4}[][]{$r_{4}$}
\psfrag{i3}[][]{$r_{3}$}
\psfrag{o5}[][]{$o_{5}$}
\psfrag{nb2}[][]{$nb_{2}$}
\psfrag{nb3}[][]{$nb_{3}$}
\includegraphics[width=0.3\textwidth]{semiregular}

\end{center}
\end{figure}
%}}}
\fi

The first strategy is to enforce a sort of regularity in the prefix graphs. 
For instance, regular adders, such as Sklansky, Brent-Kung, have the inherent property that the consecutive input nodes (even and odd) are combined to create the prefix nodes at the logic level $1$. 
In our approach, we constrain this regularity for those prefix nodes (logic level $1$). 
To explain this, let us consider \Cref{fig:semiregular}. 
We can see that prefix nodes $r_1$, $r_2$, $r_3$ and $r_4$ are constructed by consecutive even-odd nodes. 
For instance, $i_{0}$ and $i_1$ are used to construct $r_1$.
But with this structural constraint, we are not allowed to construct any node by combining $i_1$ and $i_2$ as done in Kogge-Stone adders.
Note that the sub-structure, as shown in \Cref{fig:semiregular}, is a part of some regular adders like Sklansky adder,
and is imposed in our prefix structure enumeration. 
%However, unlike Sklansky adders, we allow irregularity at all logic levels except $level=1$. 

We have run experiments with $16$ bit adders, and observed that this pruning strategy (i) does not degrade the solution quality (or size of the prefix graph under same $L$ and $mfo$), 
but (ii) able to reduce the search space significantly, in comparison to not using this pruning strategy.
%However, imposing similar regularity for $level > 1$ {leads} to sub-optimal solutions for higher bit adders, such as $n \geq 16$.     

\subsubsection{Level Restriction in Non-trivial Fan-in}
Each of the prefix node $N$ ($a$:$b$), where $a$ is the most-significant-bit (MSB) and $b$ is the least-significant-bit (LSB), 
is constructed by connecting the trivial fan-in $N_{tr}$ ($a$:$c$) having same MSB as $N$, and the non-trivial fan-in $N_{non-tr}$ ($c-1$:$b$). 
For instance, in \Cref{fig:prelim}, $o_3$ and $b_2$ are respectively the non-trivial fan-in and the trivial fan-in node for the prefix node $o_5$. 
In the bottom-up enumeration technique, we put another additional restriction that the level of the trivial fan-in node is always less or equal to that of the non-trivial fan-in node, \textit{i.e.}, $\text{level}(N_{tr}) \leq \text{level}(N_{non-tr})$. 
Note that this sort of structural restriction is also inherent in regular adders, such as Sklansky or Brent-Kung adders.

In a nutshell, our PGG algorithm is a blend of regular adders and \cite{ADDER_TCAD2014_Roy}. 
We borrow some properties of regular adders to enforce in \cite{ADDER_TCAD2014_Roy} for reducing its huge search space without hampering the solution quality. 
To illustrate this, we have obtained the binary for \cite{ADDER_TCAD2014_Roy} from the authors, and first compared our result for lower bit adders, such as $n=16,32$. 
We got the solutions with same minimum size, which \textit{proves that our structural constraints have not degraded the solution quality}. 
However, for higher bit adders, we get better solution quality than \cite{ADDER_TCAD2014_Roy} as shown in \Cref{tab:synthesis}. 

\begin{table}[bt!]
\centering
\caption{Comparison with \cite{ADDER_TCAD2014_Roy} for $64$ bit adders}
\begin{tabular}{|c|c|c|c|c|}
  \hline
   \multirow{2}{*}{$mfo$} &  \multicolumn{2}{c|}{Our Approach} & \multicolumn{2}{c|}{Approach in \cite{ADDER_TCAD2014_Roy}} \\
  \cline{2-5}
  & size & Run-time (s) & size & Run-time (s) \\
  \hline \hline
  4  & 244 & 302  & 252 & 241 \\
  6  & 233 & 264  & 238 & 212 \\
  8  & 222 & 423  & - & - \\
  12 & 201 & 193  & - & - \\
  16 & 191 & 73   & 192 & 149  \\
  32 & 185 & 0.04 & 185 & 0.04 \\
  \hline
\end{tabular}
\label{tab:synthesis}
\end{table} 

Column $1$ presents the $mfo$ constraint, while columns $2$ and $3$ respectively show the size and run-time for our enhanced algorithm, 
and the corresponding entries for \cite{ADDER_TCAD2014_Roy} are respectively represented in columns $4$ and $5$. 
In general, when fan-out is relaxed or $mfo$ is higher, the run-time is less due to relaxed size-pruning as explained in \cite{ADDER_TCAD2014_Roy}.
Note that \cite{ADDER_TCAD2014_Roy} cannot generate solutions for $mfo=8,12$ due to generation of innumerable intermediate solutions as explained in \cite{ADDER_ASPDAC2015_Roy}.
On the contrary, our structural constraints can do a pre-filtering of the potentially futile solutions, 
thereby allowing relaxed size-pruning and size-bucketing to search for more effective solution space. 
In terms of run-time, it is slightly worse in a few cases, but importantly, this generation is a one-time process, 
and this run-time is negligible in comparison to the design space exploration by the physical design tools. 
So our imposed structural restrictions (i) do not degrade the solution quality, 
(ii) achieve better solution sizes for all $mfo$ than \cite{ADDER_TCAD2014_Roy} for higher bit adders which could not even generate solutions in all cases, 
and (iii) help to obtain wider physical solution space to be demonstrated in Section \ref{sec:wider}.   

\subsection{Quasi-random Sampling}
\label{sec:dataGen}
\iffalse
Since we can not afford to run the physical design flow for too many architectures,
and too few training data may degrade the model accuracy significantly, 
a set of adders need to be selected to represent the entire design solution space.
\fi
%One of the most intuitive way to pick the training data for supervised learning is to do a random sampling, but that may not be well representative.
%Considering the observed correlation between architectural and physical solution space (Section \ref{sec:gap}),
%we have performed an architecture driven \textbf{quasi-random} data sampling to capture the wide design space of prefix adders as described below.
We have mainly focused on $64$ bit adders in this work as this is mostly used in today's microprocessors. 
%In order to encapsulate the wide design space of adders, we generate around $25$k prefix architectures with max fan-out $mfo=4,6,8,12,16,20,24,28,32$. 
From all prefix adder solutions, 
%we sample around $300$ solutions via the quasi-random approach which is conducted by a two-level binning ($mfo$, $s$) followed by random selection, 
%which are used for training ($250$) and testing ($50$).  
we sample a set of solutions for building the learning model via the quasi-random approach which is conducted by a two-level binning ($mfo$, $s$) followed by random selection, 
%This approach aims to evenly sample the prefix adders covering different architectural bins.
This approach aims to evenly sample the prefix adders covering different architectural bins.
The primary level of binning is determined by $mfo$ of the solutions. 
However, there may be thousands of architectures sharing the same $mfo$, so the secondary level of binning is based on $s$.
Afterwards, adders are picked randomly from those secondary bins. 

We illustrate the quasi-random sampling with the following example: given $5000$ solutions with $mfo=4$, we want to pick $50$ solutions from them. 
Suppose these $5000$ solutions have the size distribution from $244$ to $258$. 
First a random solution is picked from the bucket of the solutions ($mfo=4$, $s=244$).
Then we pick a solution randomly from ($mfo=4$, $s=245$), and so on. 
After picking $15$ solutions from each of those buckets with $mfo=4$, we again start from the bucket ($mfo=4$, $s=244$). 
This process is repeated until we get $50$ solutions. Similar procedure is done with other $mfo$ values.

\subsection{Physical Solution Space Comparison with ~\cite{ADDER_TCAD2014_Roy}}
\label{sec:wider}
\begin{figure}[tb!]
	\centering    \hspace{-.2in}
	\subfloat[]{\includegraphics[width=.25\textwidth]{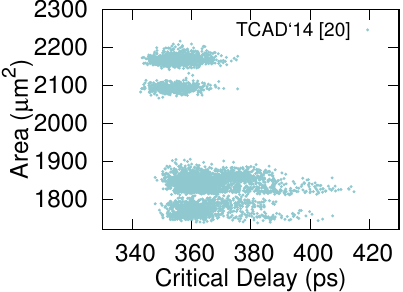}    \label{fig:wider-ad-1}} %\hspace{.2in}
	\subfloat[]{\includegraphics[width=.25\textwidth]{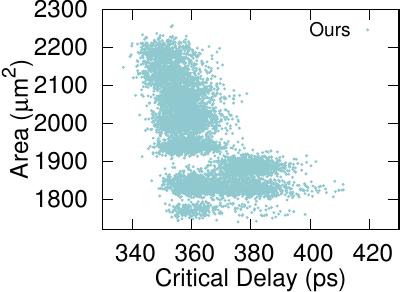} \label{fig:wider-ad-3}}  \hspace{.1in}
	\subfloat[]{\includegraphics[width=.25\textwidth]{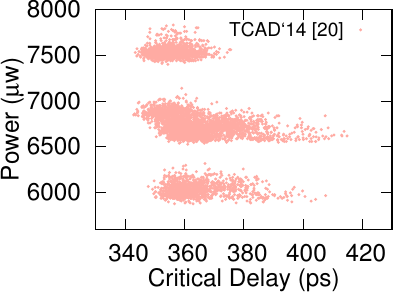}    \label{fig:wider-pd-1}} %\hspace{.2in}
	\subfloat[]{\includegraphics[width=.25\textwidth]{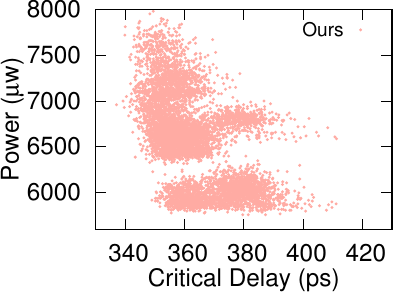} \label{fig:wider-pd-3}}  \hspace{.1in}
	\caption{
			Quasi-random sampled adders vs.~adders from \cite{ADDER_TCAD2014_Roy}.
			(a) Solution space in area vs.~delay domain from \cite{ADDER_TCAD2014_Roy};
			(b) Solution space in area vs.~delay domain from ours;
			(c) Solution space in power vs.~delay domain from \cite{ADDER_TCAD2014_Roy};
			(d) Solution space in power vs.~delay domain from ours.
	}
	\label{fig:wider}
\end{figure}

In this subsection, we show the usefulness of our algorithm for obtaining wider solution space in physical design domain in comparison to \cite{ADDER_TCAD2014_Roy}.
%Our adder synthesis algorithm is an enhancement to~\cite{ADDER_TCAD2014_Roy}. %and falls into the second category.
%For comparing with~\cite{ADDER_TCAD2014_Roy},
%we have obtained the binary for~\cite{ADDER_TCAD2014_Roy} from the authors and 
Among the prefix adders generated by ~\cite{ADDER_TCAD2014_Roy}, we randomly sampled 7000 prefix adders.
Those prefix adders are fed into the full EDA flow (synthesis, placement and routing) to get their real delay, power and area values (takes around 700 hours).
We plot these adders by~\cite{ADDER_TCAD2014_Roy} and our representative $3000$ adders in \Cref{fig:wider}.
It can be seen that, although the numbers of adders by~\cite{ADDER_TCAD2014_Roy} is more than $2$ times of our representative adders, 
our adders still cover \textit{wider} solution space in physical domain, 
demonstrating the effectiveness of our enhanced algorithm PGG.
This is in accordance with the solutions missed by~\cite{ADDER_TCAD2014_Roy} as mentioned in \Cref{tab:synthesis}.
Those availabilities eventually offer more opportunities for our machine learning methodology to identify close to ground truth Pareto frontier solutions.

\iffalse
%{{{
\newcommand{\twopartdef}[4]
{
	\left\{
		\begin{array}{ll}
			#1 & \mbox{if } #2 \\
			#3 & \mbox{} #4
		\end{array}
	\right.
}
%}}}
\fi

\section{Bridging Architectural Solution Space to Physical Solution Space}
\label{sec:mladder}
In most EDA problems, the metrics of the solution quality are typically conflicting.
For instance, if we optimize the timing of the design, then the power/area may be compromised and vice versa.
So one imperative job of EDA engineers is to find the Pareto-optimal points of the design
enabling the designers to select among those.
In this section, we first provide the preliminaries about Pareto optimality, and the error metrics of Pareto optimal solutions.
Then we discuss the gap between the prefix architectural solution space and physical solution space in adders,
which motivates the need of the machine learning-based approach for optimal adder exploration. 
Finally, a domain knowledge-based feature selection details are presented along with training data sampling for the learning models.
\subsection{Preliminaries}
\label{sec:pal-prelim}
\begin{mydefinition}[Pareto Optimality]
	An objective vector $\mathbf{f}(\mathbf{x})$ is said to dominate $\mathbf{f}(\mathbf{x}')$ if:
	\begin{equation}
	\begin{aligned}
	&\forall i \in [1,n], f_i(\mathbf{x}) \leq f_i(\mathbf{x}') \\ 
	\textrm{and } &\exists j \in [1,n], f_j(\mathbf{x}) < f_j(\mathbf{x}').
	\end{aligned}
	\end{equation}
\end{mydefinition}

A point x is \emph{Pareto-optimal} if there is no other $\mathbf{x}'$ in design space such that
$\mathbf{f}(\mathbf{x}')$ \emph{dominates} $\mathbf{f}(\mathbf{x})$.

As in this paper for adder design, a Pareto-optimal design is where none of the objective metrics, such as area, power or delay, can be improved without worsening at least one of the others. 
The \emph{Pareto Frontier} is the set of all the Pareto-optimal designs in the \emph{objective space}.
Therefore, the goal is to identify the Pareto-optimal set $P$ for all the Pareto-optimal designs.

\begin{figure}[tb!]
	\centering
	\includegraphics[width=0.36\textwidth]{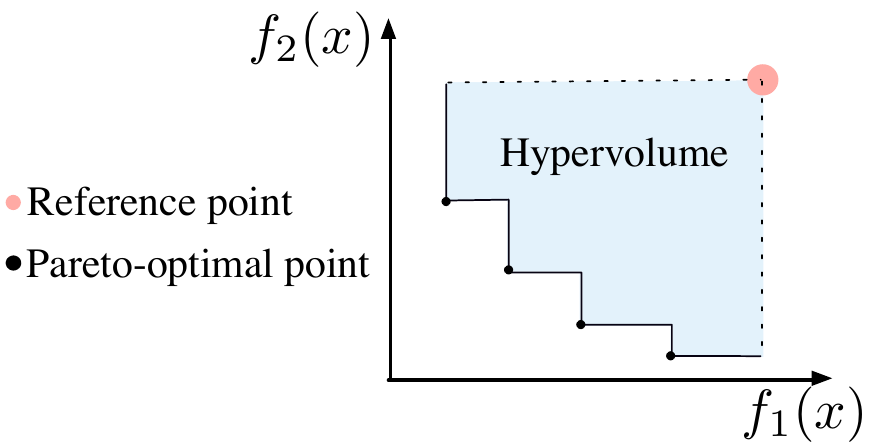}
    \caption{Hypervolume with two objectives in objective space.}
	\label{fig:Hvol}
\end{figure}

\begin{figure}[tb!]
    \centering
    \subfloat[]{\includegraphics[height=3.0cm]{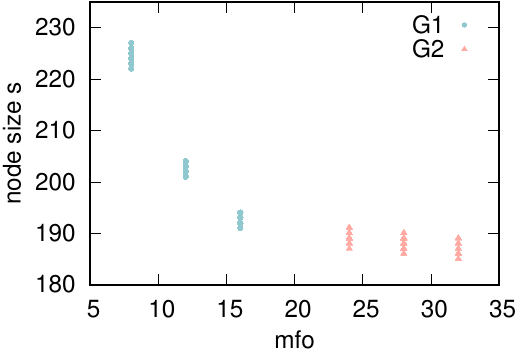}     \label{fig:prefixSpace}}
    \subfloat[]{\includegraphics[height=3.0cm]{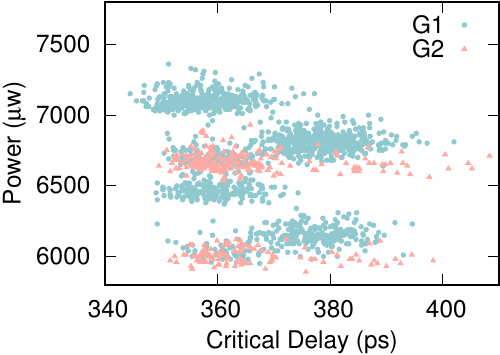}     \label{fig:phySolnSpace}}
    \caption{Gap between prefix structure and physical design of adders:
        (a) Architectural solution space;
        (b) Physical solution space.
    }
    \label{fig:gap}
\end{figure}

\begin{figure*}[tb!]
    \centering
    \begin{minipage}{.72\linewidth}
        \subfloat[]{\includegraphics[height=3.3cm]{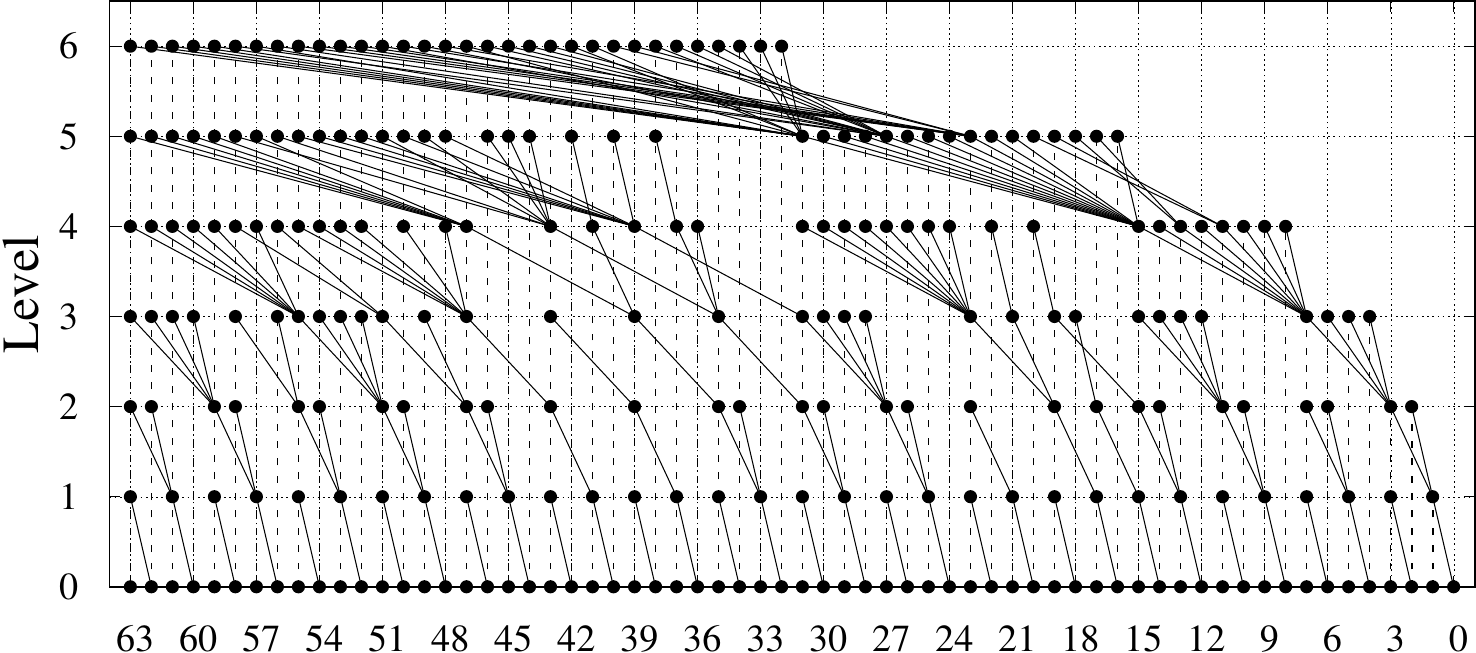}  \label{fig:gap-a}}
        \hspace{.2in}
        \subfloat[]{\includegraphics[height=3.3cm]{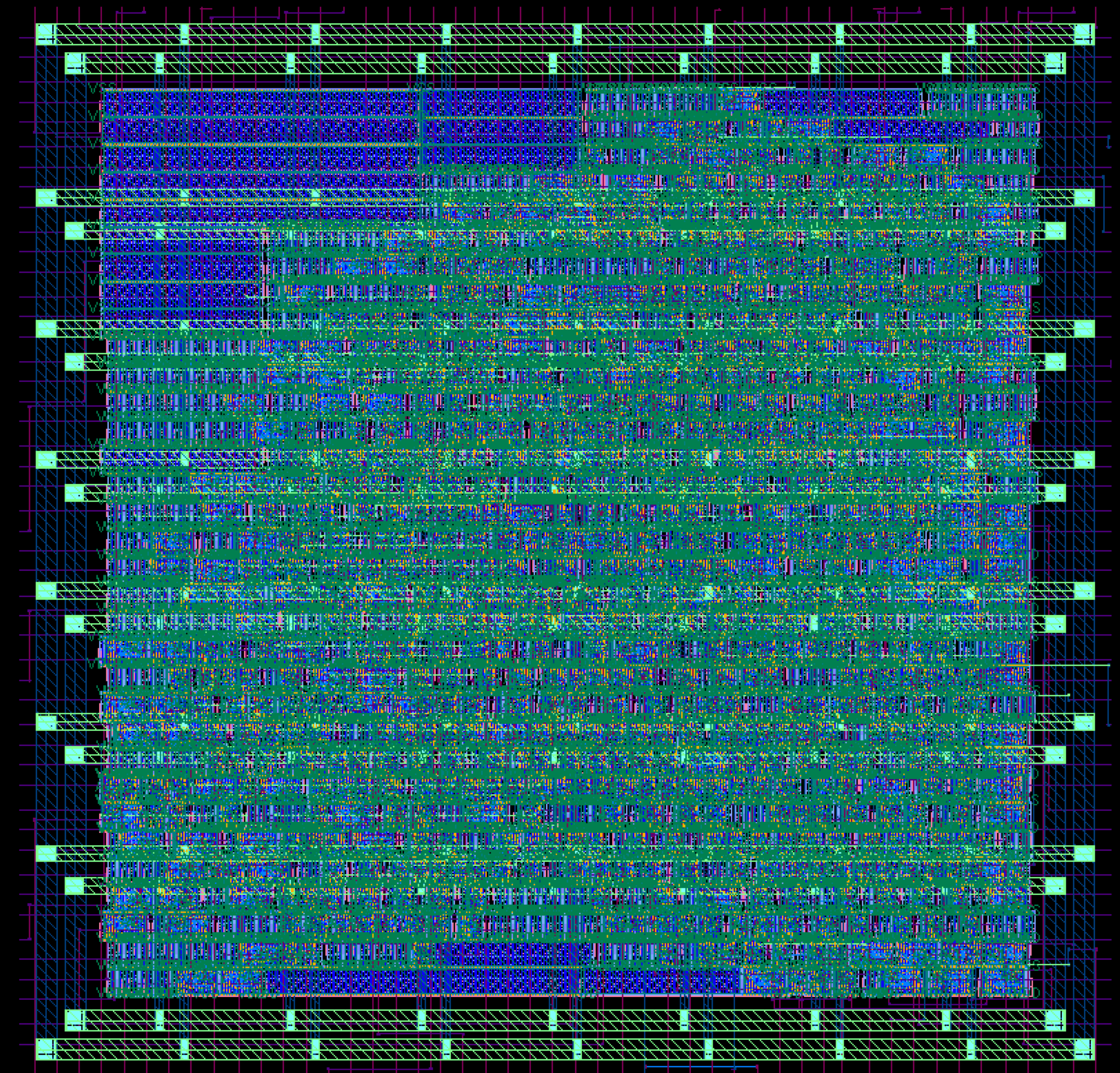}  \label{fig:gap-b}}
        \caption{
            (a) An example of architectural solution: Bit-width = 64, size = 201, Max.~level \\
            = 6, Max.~fanout = 12; (b) Corresponding physical solution.
        }
        \label{fig:gap-ex}
    \end{minipage}
    \begin{minipage}{.24\linewidth}
        \includegraphics[height=3.08cm]{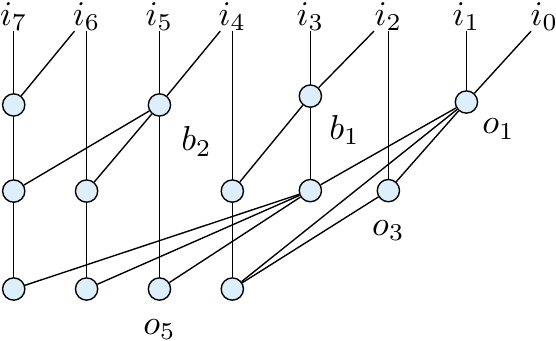}
        \caption{Defining $spfo$ of a node.}
        \label{spfo_definition}
    \end{minipage}
\end{figure*}

\begin{mydefinition}[Hypervolume]
	The hypervolume computes the volume enclosed by the Pareto frontier and the reference point in the objective space  \cite{LEARN_EMO2007_Zitzler}.
\end{mydefinition}
%To evaluate the prediction quality, \emph{hypervolume} \cite{LEARN_EMO2007_Zitzler} is introduced which computes the volumes enclosed by the \emph{Pareto frontier} and the reference point. 
In \Cref{fig:Hvol}, the shaded area is an example of the hypervolume of a Pareto set with two objectives. 
Then the \emph{hypervolume error} for a predicted Pareto set $\hat{P}$ is defined as
\begin{equation}
\eta = \frac{V(P)-V(\hat{P})}{V(P)},
\end{equation}
where $P$ is the true Pareto-optimal set,
and $V(P)$ is the hypervolume of the Pareto set $P$.
Note that a prediction $\hat{P}$ which contains the whole design space has an error of 0.
Thus the predicted set $\hat{P}$ with less points is desired.
%The trivial prediction $\hat{P}$ contained the whole design space has error 0; thus,
%a significant point of a reasonable algorithm is that the predicted $\hat{P}$ contains
%few or no \emph{dominated} points at all.

\iffalse
In the following subsections, we discuss the gap between the prefix architectural solution space and physical design solution space in adders,
which motivates the need of the machine learning based approach for optimal adder exploration. 
Since features are essential for building the learning model, 
a domain-knowledge based feature selection details are presented.
\fi

%We then present the quasi-random data sampling for our learning models.
%Finally, we propose our learning based Pareto frontier exploration methodology for prefix adder architectures, with trade-off among delay, power and area.
%to tackle the intractable runtime that would have been resulted by exhaustively running physical design flow for all individual prefix network solutions.

\subsection{Gap Between Logic and Physical Design} 
\label{sec:gap}

Since we focus on high performance adders and explore the prefix adders of logic level $L=\log_2 n$, the metrics at this architecture stage are prefix node size $s$ and max fan-out $mfo$. 
These two metrics are conflicting, \textit{i.e.}, if we reduce $mfo$, $s$ increases and vice-versa. 
Similar competing relationship exists between delay and power/area after physical design. 
It should be stressed that power and $s$ are correlated, and $mfo$ indirectly controls the timing as more restricted fan-out can mitigate congestion and load-distribution, thereby improving the delay of the adder. 
However, this relationship between architectural synthesis and physical design is approximate, and not a very high-fidelity one.

To demonstrate this, we plot node size $s$ vs.~$mfo$ and power vs.~delay in \Cref{fig:gap} for several $64$ bit adder solutions.
In this experiment, we have generated the prefix architecture solutions by PGG, and the final power/delay numbers are obtained by running those solutions through EDA tools as explained later in Section~\ref{sec:result}.
An example of the prefix architecture and the corresponding physical solution is presented in \Cref{fig:gap-ex}.
In \Cref{fig:prefixSpace}, we broadly categorize the solutions into $2$ groups, (i) $G_{1}$ with higher node size and lower $mfo$, and (ii) $G_{2}$ with lower node size and higher $mfo$.
In \Cref{fig:phySolnSpace}, the same designs as \Cref{fig:prefixSpace} are projected into the physical solution space, restoring the group information.
Design Compiler \cite{TOOL_dc} (version F-2011.09-SP3) is used for logical synthesis, and IC Compiler \cite{TOOL_icc} (version J-2014.09-SP5-3) is used for the placement and routing.
Non linear delay model (NLDM) in 32$nm$ SAED cell-library \cite{LIB_SAED} is used for technology mapping.
The key observations here are firstly, \textit{there is a correlation between architectural solution space and physical design solution space}.
For instance, the solutions from $G_{1}$ are mostly on the upper side, and those of $G_{2}$ are mostly on the lower side in \Cref{fig:phySolnSpace},
thereby indicating a correspondence between $s$ and power. 
Nevertheless, \textit{it is not completely reliable}. 
For example, (i) the delay numbers for $G_{1}$ and $G_{2}$ are very much spread,
(ii) a cluster can be observed where the solutions from $G_{1}$ and $G_{2}$ are mixed up in \Cref{fig:phySolnSpace}, 
and (iii) several solutions of $G_{1}$ are better than several solutions of $G_{2}$ in power, which is not in accordance with the metrics at the prefix adder architecture stage. 
So we can not utterly rely on architectural solution space to achieve the optimal output in physical solution space.

However, since our algorithm generates \textbf{hundreds of thousands} of prefix graph structures,
it is intractable to run synthesis and physical design flows for even a small percentage of all available prefix adder architectures.
To address this \textbf{fidelity gap} between the two design stages and the high computational cost together,
we come up with a novel machine learning guided design space exploration as replacement of exhaustive search.

\iffalse
%{{{
\begin{figure}[htb!]
    \centering
    \subfloat[]{\includegraphics[width=.48\textwidth]{fig1a}}
    \hspace{.1in}
    \subfloat[]{\includegraphics[width=.48\textwidth]{fig1b}}
    \caption{}
\end{figure}
%}}}
\fi

%The reason for this quasi-random selection is to ensure that the training/testing data will capture the wide-design space of prefix adders. 

\subsection{Feature Selection} 
\label{sec:feature}

The feature is a representation which is extracted from the original input representation, and it plays an important role in machine learning tasks.
We now discuss the features to be used for the learning model.
Features are considered from both prefix adder structure and tool settings, with a focus on the former.
We select node size and maximum-fan-out ($mfo$) of a prefix adder as two main features for our learning model.
However, for any given $mfo$ and node size, there will be hundreds or even thousands of different prefix architectures.
Therefore, additional features are required to better distinguish individual prefix adder attributes.
We define a parameter sum-path-fan-out ($spfo$) for this. 
Let $a$ and $b$ are the fan-in nodes of a node $n$, then $spfo(n)$ is defined recursively as:
\begin{equation}
    spfo(n) = \twopartdef {0,} {n \in \text{input},} {\text{sum}(fo(a) + spfo(a),\\ \quad \quad fo(b) + spfo(b)),}{\text{otherwise}.}
\end{equation}

Here $fo(n)$ denotes the fan-out of any node $n$.
Consider the prefix adder structure in \Cref{spfo_definition}, and according to the definition we have:
\begin{align*}
    spfo(o_{1}) &= \text{sum}(fo(i_{0}) + spfo(i_{0}), fo(i_{1}) + spfo(i_{1})) \\
                &= \text{sum}(1,1) = 2,\\
    spfo(b_{1}) &= \text{sum}(fo(i_{2}) + spfo(i_{2}), fo(i_{3}) + spfo(i_{3}))\\
                &= \text{sum}(2,1) = 3,\\
    spfo(b_{2}) &= \text{sum}(fo(i_{4}) + spfo(i_{4}), fo(i_{5}) + spfo(i_{5}))\\
                &= \text{sum}(2,1) = 3.
\end{align*}
Therefore, we can use the recursive definition to calculate
\begin{align*}
    spfo(o_{3}) &= \text{sum}(fo(o_{1}) + spfo(o_{1}), fo(b_{1}) + spfo(b_{1}))\\
                &= \text{sum}(3+2,2+3) = 10,\\
    spfo(o_{5}) &= \text{sum}(fo(o_{3}) + spfo(o_{3}), fo(b_{2}) + spfo(b_{2}))\\
                &= \text{sum}(3+10,3+3) = 19.
\end{align*}
%The $spfo$ of any node can also be defined similar to $mpfo$ with just replacing `max' operator with `sum' operator. 
%We omit the detail illustration for $spfo$ due to space constraint. 

\iffalse
%{{{
\begin{figure}[tb!]
\begin{center}
\psfrag{x0}[][]{$i_{0}$}
\psfrag{x1}[][]{$i_{1}$}
\psfrag{x2}[][]{$i_{2}$}
\psfrag{x3}[][]{$i_{3}$}
\psfrag{x4}[][]{$i_{4}$}
\psfrag{x5}[][]{$i_{5}$}
\psfrag{x6}[][]{$i_{6}$}
\psfrag{x7}[][]{$i_{7}$}
\psfrag{b1}[][]{$b_{1}$}
\psfrag{b2}[][]{$b_{2}$}
\psfrag{o1}[][]{$o_{1}$}
\psfrag{o3}[][]{$o_{3}$}
\psfrag{o5}[][]{$o_{5}$}
\psfrag{nb2}[][]{$nb_{2}$}
\psfrag{nb3}[][]{$nb_{3}$}
\includegraphics[width=0.25\textwidth]{prefix_structure}
%\vspace{-10pt}
\end{center}
\end{figure}
%}}}
\fi

In our methodology, we use the $spfo$ of the output nodes which are at $\log_2 n$ level (there are $32$ nodes at level $6$ for $64$ bit adder) as the features to characterize the prefix structures, in addition to $mfo$, size and target delay. 
The basic intuition for selecting $spfo$ of the output nodes as the features is that the critical path delay of the adder is the longest path delay from input to output. So it depends on the (i) path-lengths, 
which can be represented at the prefix graph stage by the logic level of the node, 
and (ii) the number of fan-outs driven at every node on the path. 
Note that we have skipped the $spfo$ of the output nodes which are not at $\log_2 n$ level as for those nodes, the path length is smaller, and those would not potentially dictate the critical path delay.

Apart from these prefix graph structural features, we also consider tool settings from synthesis stage and physical design stage as other features. 
We have synthesized the adder structures using industry-standard EDA synthesis tool \cite{TOOL_dc}, where we can specify the target-delay for the adder. 
The tool then adopts different strategies internally to meet that target-delay which we can hardly take into account during prefix graph synthesis. 
Consequently, changing the target-delay can lead to different power/timing/area metrics. 
So we have considered target-delay as a feature in our learning approach.

%We also consider the tool settings in physical synthesis stage.
In physical design, utilization is an important parameter, which defines the area occupied by standard cell, macros and blockages.
Different utilization values can lead to different layouts after physical design.
Therefore, we take utilization as another feature in the learning model.

In addition to the target delay and utilization, other tool settings have also been explored.
The optimization level setting in logical synthesis has a potential impact on the performance of adders,
which can be adjusted by \texttt{compile} and \texttt{compile\_ultra} commands with different options.
After synthesizing, it is observed that the solutions generated with \texttt{compile\_ultra} can significantly dominate the solutions generated by \texttt{compile}.
Therefore, this setting is fixed to \texttt{compile\_ultra} level as we are aiming at superior designs.

In this work, the technology node is not used as a feature.
From the machine learning perspective, there is a common assumption for conventional machine learning applications that the training and test data are drawn from the same feature space and the same distribution \cite{LEARN_TKDE2010_Pan}.
The values of area/power/delay may vary a lot under different technology nodes, which results in different underlying data distributions. 
Therefore, the technology node for synthesis should be consistent.
The proposed approach for feature extraction can also be applied to other technology nodes as long as the technology node is consistent during the design flow.
If the technology node of the testing data switches to another one, the machine learning model should be re-trained using the data from that technology node to ensure the accuracy of the model.

\subsection{Data Sampling}
Since we can not afford to run the physical design flow for too many architectures,
and too few training data may degrade the model accuracy significantly, 
a set of adders need to be selected to represent the entire design solution space.
However, finding a succinct set of representative training data for the traditional supervised learning is difficult.
In order to tackle this difficulty, we come up with two learning approaches in the next two sections.
The first one is the passive supervised learning where a quasi-random data sampling is performed to obtain the training data,
followed by multi-objective scalarization to achieve the Pareto optimal solutions.
The second one is the active learning approach where model training is integrated to finding Pareto-optimal frontiers of the design space.

\section{$\alpha$-sweep learning}
\label{sec:psl}
\begin{figure}
	\centering
	\includegraphics[width=.82\linewidth]{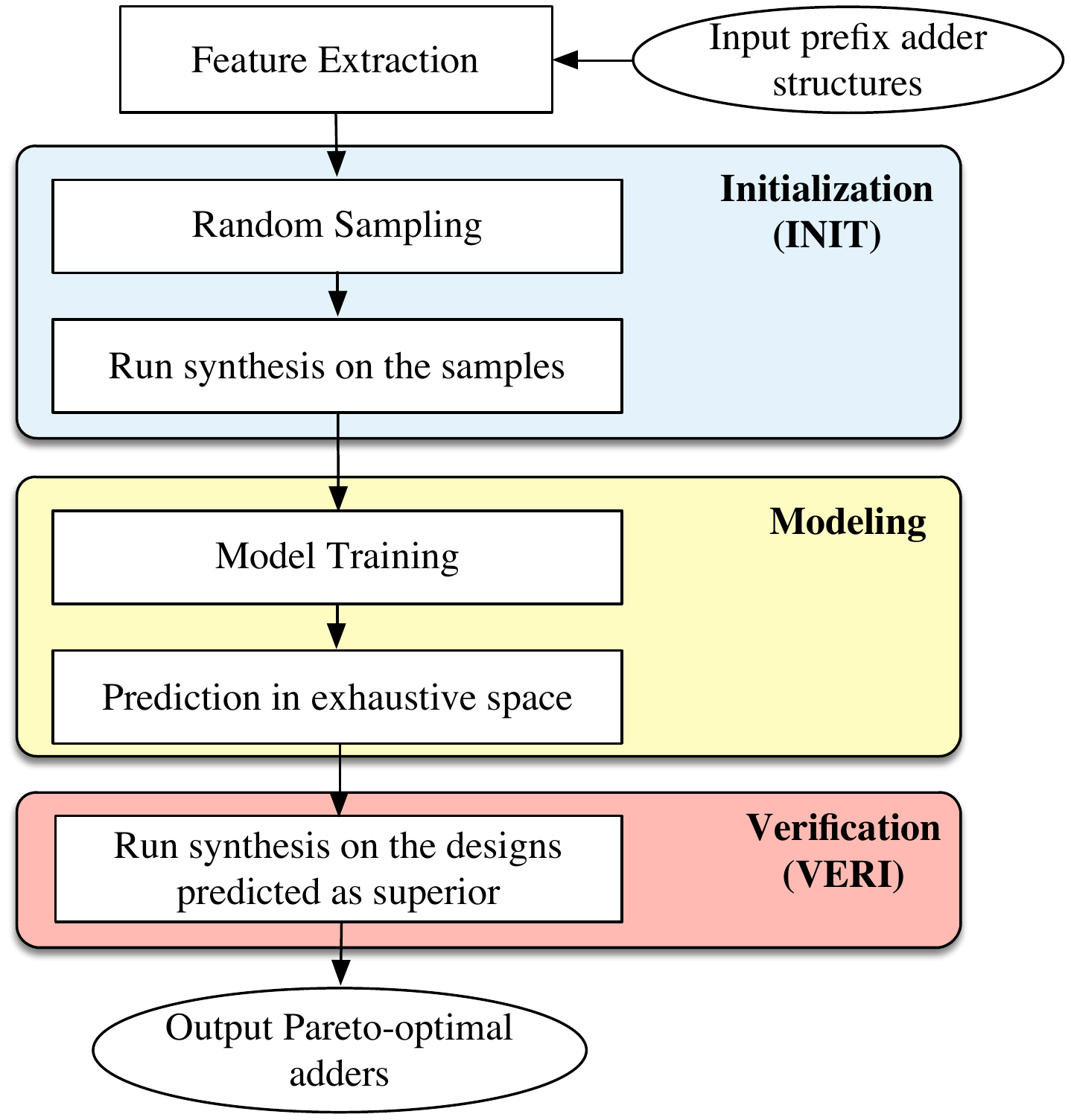}
    \caption{Overall flow of $\alpha$-sweep learning.}
	\label{fig:flow-alpha}
\end{figure}

%We introduce two learning approaches in this section, including passive supervised learning and Pareto active learning.
In this section we propose a pareto-frontier exploration flow which is based on support vector machine. 
The overall flow of our $\alpha$-sweep supervised learning-based Pareto-frontier exploration is presented in \Cref{fig:flow-alpha}.

\subsection{Scalarization to the Single-Objective}
\label{sec:sinobj}

In this work, supervised learning is preferred over unsupervised learning since supervised learning has a substantial advantage over unsupervised learning for our problem. 
In particular, supervised learning allows to take advantage of the golden result, i.e., the true area/power/delay, generated by the synthesis tools for each design, 
instead of just letting the algorithm work out for itself what the classes should be.
In general, supervised learning usually outperforms the unsupervised learning for this kind of regression and classification tasks.

Before applying machine learning for exploring Pareto frontier, we first validate the effectiveness of the features we extract by building regression models for single metric prediction.
For learning models, we explored (i) several supervised learning techniques, such as linear regression, Lasso/Ridge, Bayesian ridge model and support vector regression (SVR) with linear, polynomial and radial-basis-function (RBF) kernel, 
and (ii) $36$ features, including $4$ primary features, size, $mfo$, target delay and utilization (tool settings), and $32$ secondary features for $spfo$. 
We observed that we could get an $R^2$ score above $0.95$ for area and power even with primary features and linear models. 
However, we don't get good scores for delay with only primary features. 
Best model fitting for delay is achieved with SVR (RBF kernel) with these $4$ primary and $32$ secondary features. 
Since SVR with RBF kernel give good MSE (mean-squared-error) scores for all metrics, delay, area and power, we have used this model throughout for design space exploration.
%In total $300$ representative adders have been used for training/testing, and increasing the training set size further did not give us any significant improvement in model accuracy.

The model experiments give us the following key insights: 
(i) \textbf{tool setting} can play an important role in building the learning models in EDA.
For instance, MSE scores for area and power improve from $0.021$ to $0.003$, and $0.228$ to $0.027$ respectively when we add the `target delay' feature in our model building, 
%(ii) $32$ $spfo$ features are more important than $32$ $mpfo$ features.
(ii) secondary features play an important role in improving the model accuracy.
For instance, when we include $spfo$ features in model building, MSE score for delay improves from $0.200$ to $0.170$. 
%But it doesn't improve further with introducing $32$ $mpfo$ features. 
%This is intuitively due to our observation that the $mpfo$ features are not unique, unlike $spfo$, 
%\textit{i.e.}, different prefix graph structures are found to have same $mpfo$ features, which is not the case for $spfo$, 
(iii) linear models are not sufficient for modeling delay. 
For instance, MSE scores of delay improve from $0.214$ to $0.170$ when we go from linear models to SVR with RBF kernel, with the same set of features.

The problem of exploring the Pareto frontier of rich prefix adder space can be approached by first sampling a subset of prefix adder architectures, 
and generating the power, area, delay numbers of each prefix adder by running through the logic synthesis and physical design flow.
Those known data set will be used as the training and testing data for supervised machine learning guided model fitting.
Once the model is fitted, we can apply the exhaustive prefix adder architectures to this model and get the predicted Pareto frontier solution set.
This is due to the merit of much faster runtime for a machine learning model in prediction stage than running the entire VLSI CAD flow.

However, conventional machine learning problem aims at maximizing the prediction accuracy rather than exploring a Pareto frontier out of a solution set.
Improving the model accuracy does not necessarily improve the Pareto frontier and the direct use of the fitted model for Pareto frontier exploration can even miss up to $60\%$ Pareto frontier points~\cite{LEARN_DATE2016_Meng}.
We therefore need a machine learning integrated Pareto frontier exploration methodology, where the Pareto frontier selection does not rely only on the model accuracy.
So we develop a fast yet effective algorithmic methodology, enabled by regression model to explore the Pareto frontier of prefix adder solutions.

%\subsubsection{Pareto-Frontier in 2D Design Space}
First we consider two spaces for Pareto frontier exploration: the delay vs.~area as well as the delay vs.~power.
For either space, there exists a strong trade-off between the two metrics.
For delay vs.~power space, we propose to use a joint output Power-Delay function ($PD$) as the regression output rather than using any single output.
\begin{equation}
PD = \alpha \cdot Power + Delay.
\label{eq:pd}
\end{equation} 

The rationale of using scalarization \cite{LEARN_ICPR1996_Tumer} or the linear summation of the power and delay metrics is that such a linear relation provides a weighted bonding between the power and the delay so that by changing the $\alpha$ value, 
the regression model will try to minimize the prediction error on the more weighted axis hence leads to more accuracy on that direction.
In contrast, the other metric direction will be predicted with less accuracy hence introducing some level of relaxations.
It can be foreseen that changing the $\alpha$ value can lead to different fitting accuracies of the regression model.
By sweeping $\alpha$ over a wide range from $0$ to large positive values, each time the regression model will be fitted to predict different best solutions which altogether form the Pareto frontier.
We call this approach $\boldsymbol{\alpha}$-\textbf{sweep}.
Note that, the $Power$ and $Delay$ values in Equation~\eqref{eq:pd} are normalized and scaled to the range between 0 and 1 by Equation~\eqref{eq:norm}.
\begin{equation}
x = \frac{x-\min(X)}{\max(X)-\min(X)}, x \in X.
\label{eq:norm}
\end{equation} 

Similarly, we have a joint output Area-Delay (AD) function for Pareto frontier exploration on Area and Delay space.
\begin{equation}
AD = \alpha \cdot Area + Delay.
\label{eq:ad}
\end{equation} 

This $\alpha$-sweep technique can be extended to simultaneously consider power, performance or delay, and area (PPA), using two scalars ($\alpha_1$ and $\alpha_2$) instead of one scalar factor $\alpha$. 
The joint output function for Pareto frontier exploration on area -- power -- delay space can be formulated as:
\begin{equation}
PPA = \alpha_1 \cdot Area  + \alpha_2 \cdot Power + Delay.
\label{eq:ppa}
\end{equation} 

The results of $\alpha$-sweep for both two-dimensional space and three-dimensional space are shown in the Section~\ref{sec:result}.
%For the sake of simplicity, we have demonstrated the methodology separately for delay vs.~area and delay vs.~power using one scalar at a time.  

%\subsubsection{Pareto-Frontier in 3D Design Space (PPA)}

%With the scalarization method, we can also perform design space exploration in 3-D space, area -- power -- delay, directly, which requires one more parameter $\beta$ to trade-off among three metrics.

\section{Pareto Active Learning}
\label{sec:pal}

In our adder design problem, obtaining the true area/power/delay values or the labeled data for each adder requires running logic synthesis and physical design flow, 
which is often time-consuming if the amount of data is huge.
Active learning is an iterative supervised learning which is able to interactively query the data pool to obtain the desired outputs at new data points.
Since the samples are selected by the learning algorithm, the number of samples to fit a model can often be much lower than the number required in traditional supervised learning. 
%This demands for an active learning approach.
Since an active sampling strategy is required in active learning, an ``uncertainty estimation'' of the prediction is needed.
Gaussian Process (GP) can make predictions and,  more importantly, provide the uncertainty estimation of its predictions by nature.
Therefore, in this paper we further propose a Pareto active learning algorithm based on Gaussian Process regression.
\iffalse
Active learning is able to interactively query the data pool to obtain the desired outputs at new data points.
Since the samples are selected by the learning algorithm, the number of samples to fit a model can often be much lower than the number required in normal supervised learning.
\fi
\subsection{Overall Flow}

\begin{figure}[tb!]
	\centering
	\includegraphics[width=.98\linewidth]{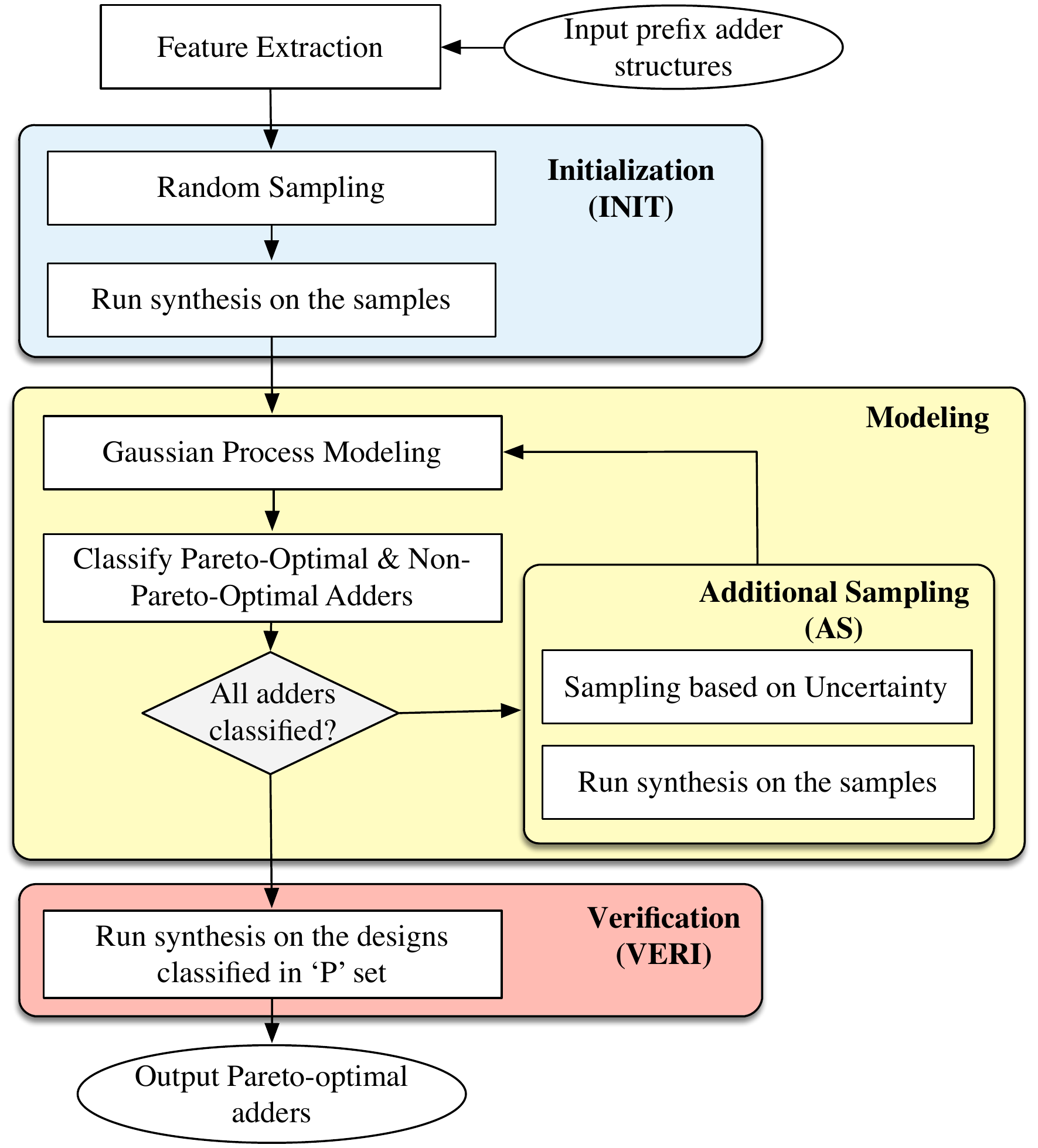}
    \caption{Overall flow of Pareto active learning.}
	\label{fig:flow-pal}
\end{figure}

The overall flow of the Pareto active learning (PAL) is shown in \Cref{fig:flow-pal}. 
Given all the prefix adder structures, first we extract the feature vector for each adder as introduced in Section~\ref{sec:feature}.
The active learning starts with Gaussian Process regression which will be illustrated later.
Unlike the passive supervised learning in which all the features and the corresponding labels are prepared in advance, 
the active learning derives the labels of each training data during the learning process on-demand.
To be specific, 
the algorithm incrementally identifies the most representative instances along with their features which are later fed into EDA synthesis flow (synthesis, placement and routing) for true area/power/delay numbers.
%the most informative instances will be selected and the corresponding metrics will be fetched. 
%Obtaining the true values for area/power/delay requires running synthesis tools with the prefix structure. 
Namely, the EDA synthesis flow and the learning process are interleaving. 
As more and more designs being selected, the model gets more and more accurate till convergence.

\subsection{Gaussian Process Prediction}
\label{sec:gpr}
A Gaussian process is specified by its \emph{mean function} and \emph{covariance function}.
A Pareto active learning scheme based on Gaussian process regression is proposed in \cite{ATL_ICML2013_Marcela}.
%The mean function $m(\mathbf{x})$ and the covariance function $k(\cdot , \cdot)$
%of a real process $f(\mathbf{x})$ are defined as
%\begin{align*}
%m(\mathbf{x})             & = \mathbb{E}[f(\mathbf{x})],\\
%k(\mathbf{x},\mathbf{x}') & = \mathbb{E}[(f(\mathbf{x}) - m(\mathbf{x}))(f(\mathbf{x}') - m(\mathbf{x}'))].
%\end{align*}
%The covariance function specifies the covariance between pairs of random variables.
%Under the guidance of prior mean and prior covariance function (kernel function),
The prior information is important to train the Gaussian Process model, which is a parameterized mean and covariance functions.
Conventionally, the training process selects the parameters in the light of training data such that the marginal likelihood is maximized.
Then the Gaussian Process model can be obtained and the regression can be proceeded with supervised input \cite{GP_B2006_Rasmussen}.
The ability of GP indicating prediction uncertainty reflects in GP learner providing a Gaussian distribution $\mathcal{N}(m(\mathbf{x}),\sigma(\mathbf{x}))$ of the values predicted for any test input $\mathbf{x}$ by computing
\begin{equation}
\hspace{-.42cm}
\begin{aligned}
m(\mathbf{x})             & = k(\mathbf{x},\mathbf{X})^{\top}(k(\mathbf{X},\mathbf{X})+\sigma^{2}\mathbf{I})^{-1}\mathbf{Y},\\
\sigma^{2}({\mathbf{x}})  & = k(\mathbf{x},\mathbf{x}) -
k(\mathbf{x},\mathbf{X})^{\top}(k(\mathbf{X},\mathbf{X})+\sigma^{2}\mathbf{I})^{-1}k(\mathbf{x},\mathbf{X}),
%\sigma^{2}({\mathbf{x}})  & = k(\mathbf{x},\mathbf{x}),
\end{aligned}
\label{eq:gp}
\end{equation}
where $\mathbf{X}$ is the training set, $\mathbf{Y}$ is the supervised information of trained set $\mathbf{X}$. 
%$\mathbf{x}'$ is any one or combination of other samples.
For Gaussian Process regression, a prediction of a design objective consists of a mean and a variance.
The mean value $m(\mathbf{x})$ represents the predicted value and the variance $\sigma(\mathbf{x})$ represents the uncertainty of the prediction.
%while the standard deviation $\sigma(\mathbf{x})$ of the distribution quantifies the confidence of the model about the prediction which implies prediction uncertainty.
%As in this paper, large standard deviation signifies the test adder design $\mathbf{x}$ is not represented by the current model under trained/sampled inputs and thus,
%contains most effective information for next training/sampling.

\subsection{Active Learning Algorithm}
\label{sec:al}
The ability of GP learners in quantifying prediction uncertainty enables a suitable application for active learning.
%The entire process is presented in Algorithm S1
%PAL algorithm with pseudo-code below is adapted from \cite{ATL_ICML2013_Marcela} to predict Pareto-optimal designs in whole adder design space with a small subset of deigns as initial input. 
Basically, three sets are maintained during the PAL process, including a set of Pareto-optimal designs ($P$), non-Pareto-optimal designs ($N$) and `unclassified' designs ($U$).

The GP models with discrepant prior are applied to learn the objective functions $f_{area}(\mathbf{x})$, $f_{power}(\mathbf{x})$, $f_{delay}(\mathbf{x})$.
%providing reference points in the architectural space that were Pareto-optimal in the physical space with high probability.
%\subsubsection{Modeling}
%Three objective functions are respectively modeled as an independent GP distribution. 
PAL calls GP inference to predict the mean vector $\mathbf{m(x)}$ and the standard deviation vector \bm{$\sigma$}$(\mathbf{x})$ of all unsampled $\mathbf{x}$ in the design space based on Equation~\eqref{eq:gp}.
Unlike other regression models such as linear regression and support vector regression, whose outputs are in form of numerical or categorical result, the output of GP is a distribution where uncertainties are involved.
To capture the prediction uncertainty for a design $\mathbf{x}$, a hyper-rectangle is defined as

\begin{equation*}
HR(\mathbf{x}) = \{\mathbf{y}: m_i(\mathbf{x})-\beta^{\frac{1}{2}}\sigma_i(\mathbf{x})\leq y_i \leq m_i(\mathbf{x})+\beta^{\frac{1}{2}}\sigma_i(\mathbf{x})\},
\end{equation*}
where $i \in \{1,2,3\}$, corresponding to area, power and delay metrics in physical space.
$\beta$ is a user-defined parameter which determines the impact of $\sigma_i(\mathbf{x})$ on the region.
%Each design $\mathbf{x}$ is assigned an uncertainty region, denoted by $U(\mathbf{x})$ which is based on the hyper-rectangle defined as
%where $i \in \{1,2,3\}$, corresponding to area, power and delay metrics in physical space.
In our implementation,  $\beta$ is set to 16 based on the analysis in \cite{ATL_ICML2013_Marcela,GP_ICML2010_Srinivas}.
%In addition, $\beta^{\frac{1}{2}}$ is scaled down by 1000 suggested by \cite{}.

As shown in \Cref{fig:flow-pal}, the PAL algorithm is an iterative process.
A few new points are selected in each iteration, and the GP model is retrained with new training set.
%In addition to modeling the uncertainty based on the output of the GP model itself,
%there are two more things which should be stressed.
%The first one is that the training set increments in each iteration, which means that the training set does not change too much in the two consecutive iterations.
Note that the model is supposed to be more and more accurate as more data being sampled.
Therefore, the uncertainty region should be smaller and smaller.
In order to ensure the non-increasing monotonicity of the uncertainty region while sampling and incorporating the previous evaluations, the uncertainty region of $\mathbf{x}$ in the $(t+1)$-th iteration is defined as 
\begin{equation}
R_{t + 1}(\textbf{x}) = R_{t}(\textbf{x}) \cap HR(\textbf{x}),
\label{eq:uncertainty}
\end{equation}
where the initial $R_0 = \mathbb{R}^n$ which is the entire objective space.

\iffalse
An example in two-dimensional space is shown in \Cref{fig:uncertainty}.
\begin{figure}
	\centering
	\includegraphics[width=.52\linewidth]{uncertainty}
    \caption{An example of uncertainty region.}
	\label{fig:uncertainty}
\end{figure}
\fi
%where $t$ starts from 0 indicating PAL iterations and $R_0 = \mathbb{R}^{3}$
%\subsubsection{Classification}
%The PAL algorithm classification is also monotonic $P$ and $N$

\begin{figure}
	\centering
	\includegraphics[width=7cm]{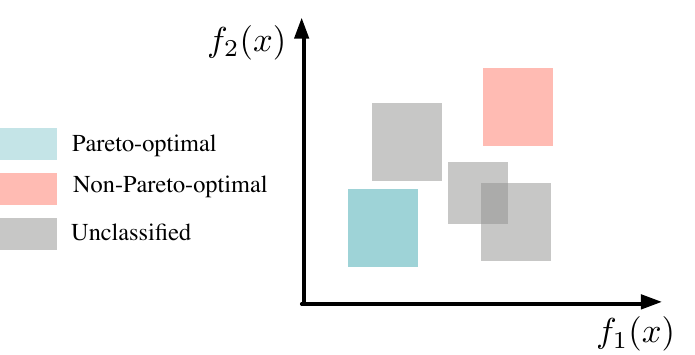}
	\caption{An example of classification.}
	\label{fig:classify}
\end{figure}

The numbers of designs in Pareto-optimal set $P$ and non-Pareto-optimal set $N$ are non decreasing as iteration $t$ increments.
Thus, at iteration $t$, the points in $P$ and $N$ keep their classification.
Intuitively, if one wants to compare the predicted performance of two designs, two extreme cases, i.e., optimistic prediction $\min(R_t(\mathbf{x}))$ and the pessimistic prediction $\max(R_t(\mathbf{x}))$ of each design, can be applied.
If the optimistic prediction of design $\mathbf{x}$ is dominated by the pessimistic prediction of other design $\mathbf{x}'$, then $\mathbf{x}$ is classified as non-Pareto-optimal;
And if the pessimistic prediction of design $\mathbf{x}$ is not dominated by optimistic prediction of any other design $\mathbf{x}'$, then $\mathbf{x}$ is classified as Pareto-optimal;
A design will remain unclassified if neither condition holds. 
\Cref{fig:classify} is presented here as an example.

In the implementation, an error tolerance $\delta$ with value $0.001$ is applied during classification.
The rules for classification can be represented as follows. 
\begin{equation}
\mathbf{x} \in \left \{
\begin{aligned}
P,  &\ \ \textrm{if} \ \ \max ( R_t (\mathbf{x})) \leq \min(R_t(\mathbf{x}')) + \delta, \\
N, &\ \ \textrm{if} \ \ \max(R_t(\mathbf{x}')) \leq \min(R_t(\mathbf{x})) + \delta, \\
U,  &\ \ \textrm{otherwise}.
\end{aligned}
\right .
\label{eq:classify}
\end{equation}

\iffalse
%The only designs $\mathbf{x}$ to be reclassified are those in $U$ (unclassified set) with tolerance error $\delta$ which is set as 0.001, is done as follows.
\begin{itemize}
	\item If pessimistic outcome $\max(R_t(\mathbf{x}))$ of $\mathbf{x}$ is not dominated by the optimistic outcome $\min(R_t(\mathbf{x}'))$ of any other design $\mathbf{x}'$, 
	%i.e., $\max ( R_t (\mathbf{x})) \leq \min(R_t(\mathbf{x}')) + \delta$ does not hold for any $\mathbf{x}' \neq \mathbf{x}$, 
	then $\mathbf{x}$ is classified as Pareto-optimal;
	
	\item If the optimistic outcome $\min(R_t(\mathbf{x}))$ of $\mathbf{x}$ is dominated by the pessimistic outcome $\max(R_t(\mathbf{x}')$ of some other design $\mathbf{x}'$,
	%\begin{equation}
	% \max(R_t(\mathbf{x}')) \leq \min(R_t(\mathbf{x})) + \delta,
	%\end{equation}
	then $\mathbf{x}$ is classified as non-Pareto-optimal;
	
	\item All other designs remain unclassified.
\end{itemize}
\fi

After classification in each iteration, a new adder design with the largest length of the diagonal of its uncertainty region
$R(\mathbf{x})$ is selected for sampling. 
The value is attached to $\mathbf{x}$ as
\begin{equation}
w_t(\mathbf{x}) = \max_{\mathbf{y, y'}\in R_t(\mathbf{x})} ||\mathbf{y} -\mathbf{y}'||_2.
\label{eq:sample}
\end{equation}
%where $\mathbf{y}_{\max}$ and $\mathbf{y}_{\min}$ is the maximum objective vector and minimum objective vector of $R(\mathbf{x})$.
Intuitively, Equation \eqref{eq:sample} picks the points which are most worthy exploring.
Afterwards, these designs are going through EDA flow to get the real area, power and delay numbers, and the GP model will hence be improved with those feedback results.

%\subsubsection{Stopping criteria}

\newcommand{\argmax}{\arg\!\max}
\begin{algorithm}[!h]
	\caption{Active Learning for Pareto-frontier Exploration}
	\label{algo:PAL}
	\begin{algorithmic}[1]
		\Require Adder architectural design space $E$, 
		GP prior, maximum iteration number $T_{\max}$;
		\Ensure predicted Pareto-optimal set $\hat{P}$;
		\State $P \gets \emptyset, N \gets \emptyset, U \gets E$; \label{pseudo:init:start}
		\State Randomly select a small subset $X = \{\mathbf{x}_i\}$ of $E$;
        \State Get true values $Y = \{\mathbf{y}_i | \mathbf{y}_i = \texttt{EDAFlow}(\mathbf{x}_i)\}$; %\Comment \emph{Initialization};
		%\State Initialize set 
		\State $S \gets X$;% \mathbf{x} \in \mathbf{X}, \mathbf{y} \in \mathbf{Y}\}$;
		\State $R_0(\mathbf{x}) \gets \mathbb{R}^n, \forall \mathbf{x}\in E$;
		\State $t \gets 0$; \label{pseudo:init:end}
		\While {$U \neq \emptyset$ and $t < T_{\max}$}
		%\State Obtain $\mathbf{m}_t(\mathbf{x})$ and $\bm{\sigma}_t(\mathbf{x}), \forall \mathbf{x} \in E$; %\Comment{Modeling}
		%\State $\mathbf{m}_t(\mathbf{x}) \gets \mathbf{y}(\mathbf{x})$ and $\bm{\sigma}_t(\mathbf{x}) \gets 0, \forall \mathbf{x} \in S$;
		\State Building GP model with $\{(\mathbf{x}_i, \mathbf{y}_i): \forall \mathbf{x}_i \in S \}$; \label{pseudo:model:start}
		\State Obtain $R_t(\mathbf{x}), \forall \mathbf{x} \in E$; \label{pseudo:model:end}
		%\State $P_t = P_{t-1}, N_t = N_{t-1}, U_t = U_{t-1}$ \Comment \emph{Classification}
		\ForAll { $\mathbf{x} \in U$}  \label{pseudo:classify:start}%\Comment{Classification}
		\If {$\mathbf{x} \textrm{ is Pareto-optimal}$ based on Equation~\eqref{eq:classify}} 
			%$\max$($R_t$(\mathbf{x})) $\times (1 - \varepsilon) \leq$ $\min$($R_t$(\mathbf{x})') $\times (1 + \varepsilon)$}
			%$\max ( R_t (\mathbf{x})) - \min(R_t(\mathbf{x})') \leq \delta $}
		%\State $P_t = P_t$ $\cup$ \{\mathbf{x}\}, $U_t = U_t$ $\setminus$ \{\mathbf{x}\}
		\State $P.\texttt{add}(\mathbf{x})$, $U.\texttt{delete}(\mathbf{x})$;
		\ElsIf {$\mathbf{x} \textrm{ is non-Pareto-optimal}$ based on Equation~\eqref{eq:classify}}
		%\State $N_t = N_t$ $\cup$ \{\mathbf{x}\}, $U_t = U_t$ $\setminus$ \{\mathbf{x}\}
		\State $N.\texttt{add}(\mathbf{x})$, $U.\texttt{delete}(\mathbf{x})$;
		\EndIf
		\EndFor \label{pseudo:classify:end}
		\State Obtain $w_t(\mathbf{x}), \forall \mathbf{x} \in (U \cup P) \setminus S$; \label{pseudo:sample:start}%\Comment{Sampling}
		\State Choose $\mathbf{x}' \gets \argmax\{w_t(\mathbf{x})\}$;
		\State $S \gets S \cup \mathbf{x}'$; \label{pseudo:sample:end}
		\State $t \gets t + 1$;
		\State Obtain new data $(\mathbf{x}', \mathbf{y}')$ by running EDA flow; \label{pseudo:query}
		\EndWhile
		\State $\hat{P} \gets P$; \label{pseudo:return}
	\end{algorithmic}
\end{algorithm}

The entire process is presented in Algorithm~\ref{algo:PAL}.
It starts with the initialization (lines~\ref*{pseudo:init:start}--\ref*{pseudo:init:end}).
In each iteration, the GP model is trained with the current training set $S$, and the uncertainty region for each design is obtained (lines~\ref*{pseudo:model:start}--\ref*{pseudo:model:end}).
Then the designs in the $U$ set are classified based on uncertainty regions and classification rules (lines~\ref*{pseudo:classify:start}--\ref*{pseudo:classify:end}).
After that, the design with the largest uncertainty is sampled and the sampling set $S$ is updated  (lines~\ref*{pseudo:sample:start}--\ref*{pseudo:sample:end}).
The newly sampled design is fed into synthesis tools to get the label which is used for training GP model in the next iteration (line~\ref*{pseudo:query}).
The learning process stops after all adder designs in architectural design space are classified.
The prediction is $\hat{P} = P$ (line~\ref*{pseudo:return}).
Suppose $T_{\max}$ is the maximum number of iterations, and $|E|$ is the size of solution set, then the complexity of Algorithm~\ref{algo:PAL} is at most $\calO(T_{\max}|E|)$, as maximum size of $U$ can be $|E|$. 
However, it should be stressed that although there are $T_{\max}|E|$ operations for PAL algorithm, the cost of each operation (which is a simple inference based on the Gaussian Process Regression model) is negligible in comparison to EDA synthesis flow run-time, and we will demonstrate later in \Cref{tab:runtime} that the total run-time of different approaches are dictated by the number of EDA synthesis flow runs needed in the respective approaches.

\definecolor{color-init}{RGB}{221,239,250}
\definecolor{color-model}{RGB}{255,253,181}
\definecolor{color-veri}{RGB}{255,171,164}

\section{Experimental Results}
\label{sec:result}

%In this section we demonstrate and evaluate the proposed machine learning based design space exploration methodology.
In this section we show the effectiveness of the proposed algorithms and methodologies. %in the physical solution space.
First we compare the physical solution space before/after applying PGG algorithm.
Then the Pareto frontier obtained by $\alpha$-sweep is presented. 
Next, we demonstrate the Pareto frontier obtained by active learning, and compare the quality of Pareto frontiers generated by two approaches.
Finally, we compare our explored optimal adders against legacy adders.

Since high performance adders are commonly used in CPU architectures which are typically $64$ bit, 
we have mainly presented the results for $64$ bit adders to demonstrate the methodology.
%$300$ solutions, sampled from all prefix adder solutions by the quasi-random approach, are used for training ($250$) and testing ($50$) the learning model.
However, the approach is very general to be used for adders of arbitrary bit-width.
The flow is implemented in {C++} and {Python} on Linux machine with $72$GB RAM and $2.8$GHz CPU.
We use Design Compiler \cite{TOOL_dc} (version F-2011.09-SP3) for logical synthesis, and IC Compiler \cite{TOOL_icc} (version J-2014.09-SP5-3) for the placement and routing.
\texttt{"tt1p05v125c"} corner and Non Linear Delay Model (NLDM) in 32$nm$ SAED cell-library for LVT class \cite{LIB_SAED} (available by University Program) is used for technology mapping.
Primary input activity of $0.1$ is used along with $1$GHz operating frequency for power estimation.
Regarding the tool settings, target delays of 0.1$ns$, 0.2$ns$, 0.3$ns$ and 0.4$ns$ are used.
Utilization values are set to 0.5, 0.6, 0.7 and 0.8.
We used Python based machine learning package scikit-learn~\cite{LEARN_JMLR2011_SKLearn} for the predictions.
Throughout our all experiments, the run time for machine learning predictions is less than a minute. 

\begin{figure}[tb!]
	\centering
	\pgfplotsset{
	width =0.32\textwidth,
	height=0.22\textwidth
}

\begin{tikzpicture}
\begin{axis}[
xlabel={Adder bit-width},
symbolic x coords={8, 16, 32, 64, 128, 256},
xtick={8, 16, 32, 64, 128, 256},
ylabel={Delay},
ytick={0, 4, 8, 12, 16},
ymin=0,
ymax=16
]
% use TeX as calculator:
%\addplot + [only marks, mark color=blue] coordinates {(8,193.5) (16,235.6) (32,292.0) (64,339.3) (128,411.0) (256,497.3)};
\addplot + [only marks, mark color=blue] coordinates {(8, 5.3) (16, 6.5) (32,8.1) (64,9.45) (128,11.4) (256,13.8)};
\addplot [domain=1:6, mark=., samples at = {8, 16, 32, 64, 128, 256}, color=black]{2*log2(x)-2};
%\addplot coordinates {(8,193.5) (16,235.6) (32,292.0) (64,339.3) (128,411.0) (256,497.3)};

\end{axis}
\end{tikzpicture}
    \caption{Delay values ($\times$ FO4 delay) of Kogge-Stone adders with various bit-width.}
	\label{fig:delay_bitwidth}
\end{figure}
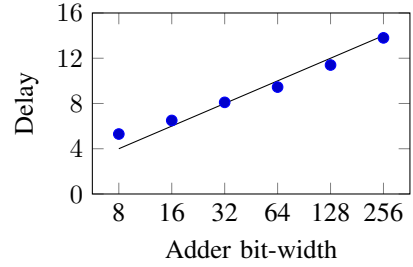

%\st{It should be stressed that SAED library used in this work may not be very realistic as that used in industry. 
%However, since this is the library available to us through university program for academic use, we have used this library.}
We relied more on the fidelity of the SAED library rather than accuracy considering that SAED library may not be very realistic as that used in industry.
For instance, the FO4 delay for a unit sized inverter for this library in the operating corner is 36$ps$ \cite{ADDER_TCAD2014_Roy,ADDER_TCAD2016_Roy}.
So 11 FO4 delay, typically being presented to be the delay for 64-bit adders in literatures \cite{ADDER_ASILOMAR2013_McAuley},
is approximately 400$ps$ which is close to the reported delays for 64-bit adders in our work.
To further demonstrate the fidelity of this library, we run the Kogge-Stone adders with bit-widths of 8, 16, 32, 64, 128 and 256 through the synthesis flow using this library. 
Then we normalize the measured delay in terms of FO4 delay, and plot it with bit-width (n) as shown in Fig.~\ref{fig:delay_bitwidth}. 
It can be seen that the delay is linear with $\log_2 n$, which is expected for a logarithmic tree adder such as Kogge-Stone adder. 
So we believe if this algorithmic methodology is applied to more realistic industrial libraries, it can show similar benefit as demonstrated with SAED $32nm$ library.  

To validate the optimality and the hypervolume error of the two learning approaches against the real world solution space,
we need to run the logical/physical EDA flow on a large set of adder solutions.
Our machine and tool set takes about 5.5 minutes to complete this full flow of a single prefix adder.
Therefore, we select a reasonable number (3000) of prefix adder solutions, which eventually took about 300 hours to complete,
but still a comparatively larger data set in comparison to our training data set. 
Crucially, those 3000 adders are also sampled in a \textbf{Quasi-random} manner in order to represent the entire solution space. 
%However, before going to the validation part, we first compare our $\alpha$-sweep with the naive single output based machine learning approach to demonstrate the effectiveness of our approach. Finally, we compare our algorithm with other adder synthesis algorithms.
%As mentioned earlier, we cannot afford to run The number of all prefix adders generated by our algorithm is in hundreds of thousands, 
%we apply the $\alpha$-sweep flow on to a \textit{representative} set of adders.

\subsection{Pareto Frontier Predicted by $\alpha$-sweep Learning}
\label{sec:naive}
%We first demonstrate the advantages of $\alpha$-sweep method over the naive single output based machine learning method.
In this experiment, we show the effectiveness of our $\alpha$-sweep learning approach. 
%We first evaluate the effectiveness of $\alpha$-sweep by comparing against the naive single output machine learning.
%For the naive machine learning, each SVR model, i.e., the model of area, power, and delay trained by the 300 adders, are independently applied onto each prefix adder solutions in the $3000$ adder set.
%From the learning predicted solutions, $50$ best delay, $50$ best area and $50$ best power solutions are picked and run through the EDA flow to generate the Pareto points for delay-area (DA) and  delay-power (DP) space. 
%Note some of the best area and best power solutions do overlap.
%The area, power, and delay values of the same adder sequence will be later associated to represent the targeted adder's prediction results.
%All adders in the delay-power (DP) and delay-area (DA) spaces form the Pareto frontiers of the DP and DA spaces, respectively.
%We selected $150$ adders sitting on or closet to the Pareto frontier of the DP and DA spaces which are fed into the standard logical and physical EDA flow to generate the real delay/area/power values.
%Eventually, the adders sitting on the Pareto frontier of the post-physical design space are plotted in \Cref{fig:naiveAlpha} by color.
%On the other hand, 
We apply the $\alpha$-sweep method with 15 different $\alpha$ values of
$(1000, 0, 100, \frac{1}{100}, 50, \frac{1}{50}, 20, \frac{1}{20}, 10, \frac{1}{10}, 8, \frac{1}{8}, 2, \frac{1}{2}, 1)$,
and collect the best $150$ solutions for delay-area and delay-power spaces where for each $\alpha$ value, the best $10$ architectures with lowest $PD$ or $AD$ values are fed into the logical/physical EDA flow to generate similar Pareto points. Note that $15+15=30$ learning models have been derived for this for all, but it is very fast as the same training data have been used, and the models are regression based. 
%If we compare two approaches in DP space, we can see in \Cref{fig:adpd:c}, the naive machine learning approach only explored the corner case of the Pareto frontier rather than the full DP space as the $\alpha$-sweep does.
%This is due to the decoupling nature of the single output based machine learning.
%Overall, this comparison illustrates the advantages of $\alpha$-sweep method.

\begin{figure}[tb!]
    \centering
    \hspace{-.2in}
    \subfloat[]{\includegraphics[width=.24\textwidth]{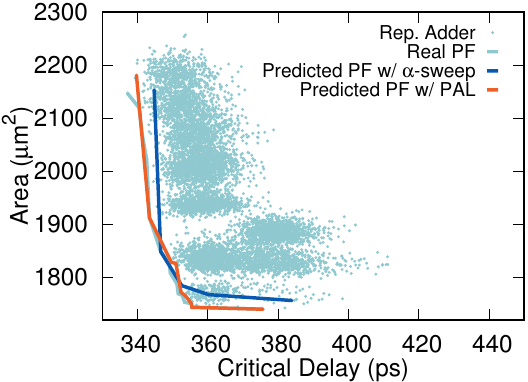}  \label{fig:ADPF-new}}
    \subfloat[]{\includegraphics[width=.24\textwidth]{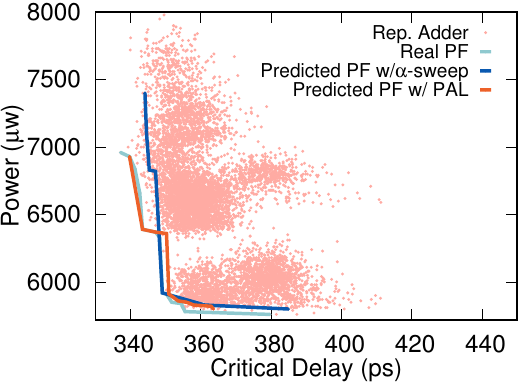}  \label{fig:PDPF-new}}
    \caption{(a) Pareto Frontier: area vs.~delay;
        (b) Pareto Frontier: power vs.~delay.}
    \label{fig:PF-compare-new}
\end{figure}

\begin{table*}[htb!]
    \caption{Comparison of different model accuracies}
    \label{tab:accuracy}
    \centering
    \resizebox{12cm}{!} {
        \begin{tabular}{|c|c|c|c|c|c|c|}
            \hline
            Model&\multicolumn{3}{c|}{MSE}  & \multicolumn{3}{c|}{Hypervolume error}  \\
            \cline{2-7}
            &Area & Power & Delay & Area-Delay & Power-Delay & Area-Power-Delay \\ 
            \hline \hline
            Original &0.003 & 0.027 & 0.170 & \textbf{0.139} & \textbf{0.122} & \textbf{0.154} \\
            Noisy    &0.024 & 0.951 & 0.711 & 0.168 & 0.148  & 0.162 \\
            \hline
        \end{tabular}
    }
\end{table*}

\Cref{fig:ADPF-new} and \Cref{fig:PDPF-new} respectively show the corresponding Pareto frontiers of the $\alpha$-sweep approach and the ground truth Pareto frontiers for the $3000$ representative adders.
Each dot in the delay-area or delay-power space indicates one adder solution after going through the logical/physical EDA flow.
We can see that generally the predicted Pareto frontier solutions are fairly close to the real Pareto frontier, with some exceptions.
Overall, the proposed approach can effectively achieve near optimal Pareto frontier without affording to spend expensive runtime on every adder.
So this learning based methodology can be readily adopted to achieve Pareto frontiers for much larger solution space which is intractable for exhaustive exploration by conventional design flow.

We have conducted additional experiments to show the impacts of the low accuracy of the machine learning model.
The basic idea is to inject random noise in the prediction stage, i.e., additional Gaussian noise is added into the predicted value.
The accuracy will be lower than original results.
Then we explore the Pareto frontier based on the noisy prediction.
Generally, the quality of the final Pareto frontier is worse than original model.
The comparison of Pareto frontier quality is presented in \Cref{tab:accuracy}.

\subsection{Comparison of the Quality of Pareto-Frontier between PAL and $\alpha$-sweep}
\label{sec:compare}

\begin{table}[tb!]
    \centering
    \caption{Pareto frontiers for PAL vs.~$\alpha$-sweep \cite{ADDER_ISLPED2017_Roy}}
    \label{tab:PF}
    \resizebox{8cm}{!}{
        \renewcommand{\arraystretch}{1.16}
        \begin{tabular}{|c|c|c|c|}
            \hline
            \multicolumn{2}{|c|}{Objective Hypervolume error}  & PAL & $\alpha$-sweep \cite{ADDER_ISLPED2017_Roy}        \\
            \hline \hline
            \multirow{2}{*}{Area-Delay} 
            & average & \textbf{0.100} &  0.139 \\
            & best    & \textbf{0.044} &  0.093\\
            \hline
            \multirow{2}{*}{Power-Delay} 
            & average & \textbf{0.109} & 0.122 \\
            & best    & \textbf{0.075} & 0.076 \\
            \hline
            \multirow{2}{*}{Area-Power-Delay} 
            & average & \textbf{0.056} & 0.154 \\
            & best    & \textbf{0.039} & 0.125 \\      
            \hline
        \end{tabular}
    }
    \begin{tablenotes}
        \small
        \item \emph{Notes:} All hypervolume error above are collected from 1000 repeated experiments.
    \end{tablenotes}
\end{table}

We implement PAL to predict Pareto-optimal designs in both two-dimensional design spaces which are area-delay space and power-delay space, as well as three-dimensional space which is area-power-delay space.
The results are compared with those of \cite{ADDER_ISLPED2017_Roy}.
The initial input set for both area-delay and power-delay is of size 250, which are randomly selected from the exhaustive design space.
The curves of Pareto frontiers for two-dimensional spaces are shown in \Cref{fig:PF-compare-new}.
%Pareto frontiers in \Cref{fig:ADPF} and \Cref{fig:PDPF} are the best performance from 1000 repeated experiments.
The hypervolume of area-delay Pareto frontiers are calculated with reference point ($\max$(delay), $\max$(area)).
%and its \emph{Hypervolume error} is $1.74 \times 10^{-2}$.
Similarly, The hypervolume of power-delay Pareto frontiers are calculated  %is 146.682 
with reference point ($\max$(delay), $\max$(power)). %and its \emph{Hypervolume error} is $9.956\times 10^{-3}$.
Note that the unit for delay is nanosecond $(ns)$ when calculating the hypervolume.
It should be stressed that there is a sort of randomness in both $\alpha$-sweep and PAL algorithm.
For $\alpha$-sweep, the training set is selected randomly.
On the contrary, the initial set in PAL is randomly selected (line 2), thereby may result in different outputs.
So \textit{the experiments are conducted for 1000 times such that the general performance is reflected}.
The comparison between two approaches are shown in \Cref{tab:PF}.
Comparing the hypervolume error of Pareto frontier obtained by PAL and $\alpha$-sweep, it can be seen that PAL achieves better performance
in predicting Pareto frontier in all design spaces, including area-delay, power-delay, and area-power-delay spaces. 

%Next we compare the total number of adders which require running EDA flow, the result of which is shown in \Cref{fig:num_samples}.
\subsection{Runtime Comparisons among Exhaustive Approach, $\alpha$-sweep and PAL} \label{sec:RTCOMP}

\begin{table*}[tb!]
	\renewcommand{\arraystretch}{1.08}
	\centering
	\caption{Comparison of runtime with single machine among different approaches} \label{tab:runtime}
    \resizebox{12.8cm}{!}{
		\begin{tabular}{|c|c|c|c|c|c|c|c|}
		\hline
            Method      & \cellcolor{color-init} \#INIT & \cellcolor{color-model}\#AS & \cellcolor{color-veri}\#VERI & \#Total & \multicolumn{3}{c|}{Runtime (mins)} \\ 
            \cline{6-8}
            & \cellcolor{color-init}& \cellcolor{color-model} & \cellcolor{color-veri}& & EDA & Modeling  & Total \\
            \hline \hline
		Exhaustive           & 10000 & -   & -      & 10000 & 55000.0 & -       & 55000.0  \\ 
		$\alpha$-sweep  & 2500   & -   & 150 & 2650   & 14575.0  & 20.0 & 14595.0  \\ 
		PAL                      &  700    & 10 & 290 & 1000   & 5500.0   & 2.0    & 5502.0   \\ \hline
		\end{tabular}
    }
    \begin{tablenotes}
	\small
	\item \emph{Notes:} The designs in ``\#INIT'' and ``\#VERI'' can be synthesized in parallel. The number of designs in each category is collected from 1000 repeated experiments.
    \end{tablenotes}
\end{table*}

There are three factors that will affect the runtime: (i) the total number of EDA synthesis runs required; (ii) Among all these required EDA synthesis runs how many of them can be parallelized; (3) The runtime of the training process in machine learning model.
All these details are recorded in \Cref{tab:runtime}.
The `INIT' represents the set of training data in the $\alpha$-sweep and the initial set in PAL, which can be parallelized because all the points are obtained in advance.
The `AS' represents the set of designs which are actively sampled during the learning process, which cannot be parallelized.
The $\alpha$-sweep approach does not involve active sampling, so the `AS' set is none here.
The `VERI' represents the set of designs which are predicted to be Pareto-optimal. 
We should run EDA synthesis flow to get the real PPA values of these designs to extract the Pareto-frontier. 
This set of designs are obtained after the learning process stops, so the EDA synthesis runs on these designs are also conducted offline, which can be parallelized. 
Each EDA synthesis run takes about 5.5 minutes.

Then we can compare the total runtime of different exploration methodologies.
For exhaustive exploration, all the prefix adders should be fed into EDA tools for synthesizing to obtain the value of each metric, which is extremely time-consuming.
There is no training, additional sampling, verification. 
The total runtime cost involves EDA flows of all the designs in the design space.
The Pareto frontier can be extracted from the results, whose runtime is much less than synthesizing and can be neglected.
The total runtime is
\begin{equation}
    T_{exh} = \frac{5.5 \times \#\textrm{INIT}}{\#\textrm{Machines}}.
\end{equation}
Since the entire solution space is so huge that one can hardly run all of them,
in our experiment, we sample representative 10K designs by random sampling.
The total runtime of synthesizing is about 55000 minutes with single machine.
It should be noted that the entire solution space is much more than 10K.

In the exploration by $\alpha$-sweep, not all adders in the design space are needed for synthesizing.
The total runtime is
\begin{equation}
    T_{\alpha} = \frac{5.5 \times (\#\textrm{INIT} + \#\textrm{VERI})}{\#\textrm{Machines}} + \textrm{Modeling time}.
\end{equation}
In our experiment, we select 2500 of the designs out of those 10K designs by random sampling to build the model, including training and testing phases.
It takes about 1.5 minutes to build the model and make predictions.
%After that, 85 designs on average in the design space are predicted to be Pareto-optimal.
%Then, we run EDA synthesis for these 85 designs to extract the Pareto-frontier.
%In total, $\alpha$-sweep method needs $2585$ designs on average when exploring in 2-D design space (area-delay and power-delay).
%The runtime cost for synthesizing is 14217.5 minutes.
When exploring in area-power-delay design space, 150 designs on average in the design space are predicted to be Pareto-optimal.
So on average 2650 designs are needed.
The runtime for synthesizing is 14575 minutes.
Note that in terms of learning models, the $\alpha$-sweep method needs to build $15 \times 15 = 225$ models since $\alpha_1$ and $\alpha_2$ both have $15$ values to choose from.

Similarly, the runtime of PAL can be calculated by 
\begin{align}
T_{PAL} &= \frac{5.5 \times (\#\textrm{INIT} + \#\textrm{VERI})}{\#\textrm{Machines}} + 5.5 \times \#\textrm{AS} \nonumber \\
   &+ \textrm{Modeling time}.
\end{align}
The size of initial set is fixed, which is 700.
It takes about 4 minutes to build the model and make predictions during the PAL process.
%When exploring the Pareto-optimal designs in $2$-D space by PAL, 7 designs on average are sampled during PAL, which cannot be synthesized in parallel.
%$74$ designs on average in the design space are predicted to be Pareto-optimal, i.e., in the $\hat{P}$ set.
%So in total $781$ designs are needed on average.
%The runtime is $4295.5$ minutes with single machine.
When exploring the Pareto-optimal designs in area-power-delay space, $10$ designs on average are sampled during PAL.
$290$ designs on average in the design space are predicted to be Pareto-optimal.
In total, $1000$ designs are needed on average.
The runtime is $5500$ minutes with single machine.
PAL algorithm needs to build $N$ models where $N$ is the number of iterations in PAL.
In our implementation, the maximum iteration is set to 20. 
It can be observed that the active learning approach outperforms the $\alpha$-sweep learning in terms of both the quality of Pareto frontier and the number of EDA flow runs. 

Note that all the runtime calculations are based on single machine.
However, the EDA synthesis runs in all three flows can be distributed to multiple machines if available, except the adders sampled during active learning, which ($10$ on average in our experiments) is very less in comparison to the total number of the synthesis runs. 
So PAL can get a significant speedup over $\alpha$-sweep and exhaustive approach with single machine and multiple machines.

%\begin{figure}[tb!]
%    \centering
%    \input{pgfplot/num_samples}
%    \caption{Comparison of number of EDA flow runs.}
%    \label{fig:num_samples}
%\end{figure}
\iffalse
\begin{figure}[tb!]
    \centering
    \input{pgfplot/runtime}
    \caption{Comparison of total runtime.}
    \label{fig:runtime}
\end{figure}
\fi

\begin{figure}[tb!]
	\centering
	\pgfplotsset{
	width =0.38\textwidth,
	height=0.26\textwidth
}

\begin{tikzpicture}
\begin{axis}[
xlabel={Number of machines},
xtick={1,...,10},
ylabel={Runtime ($\times 10^3 $ mins)},
ymin=0,
ymax=16
]
% use TeX as calculator:
\addplot [domain=1:6, mark=*, samples at = {1,...,10}, color=myblue]{5.5 * 2.650 * x^(-1)};
\addplot [domain=1:6, mark=*, samples at = {1,...,10}, color=myorange]{5.5 * 0.99 * x^(-1) + 0.01 * 5.5};
\legend{$\alpha$-sweep,PAL}
\end{axis}
\end{tikzpicture}
	\caption{Comparison of runtime with different number of machines.}
	\label{fig:rt-curve}
\end{figure}
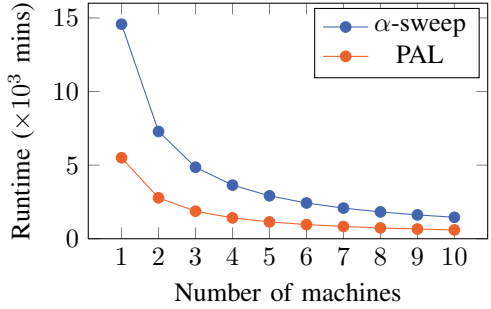

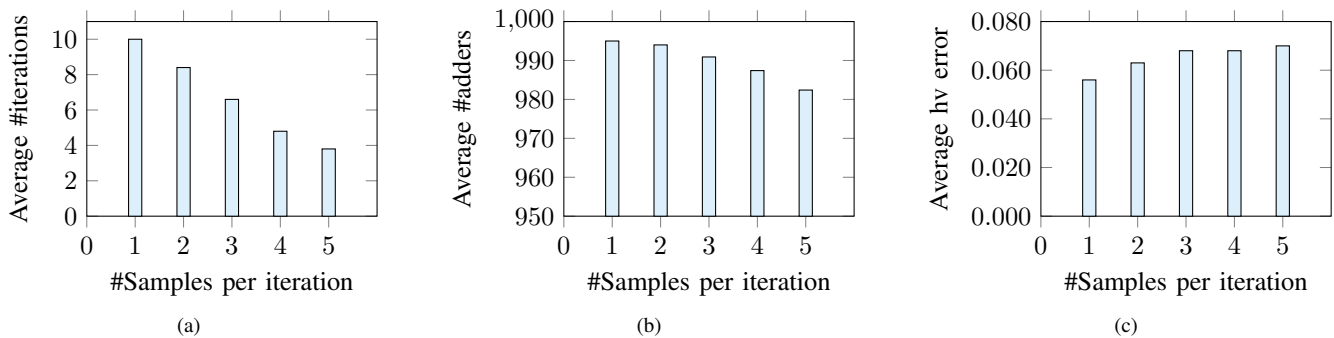
\begin{figure*}[t!]
    \centering
    \subfloat[]{\pgfplotsset{
	width =0.3\textwidth,
	height=0.23\textwidth
}

\begin{tikzpicture}[scale=1]
\begin{axis}[minor tick num=0,
ybar,
xmin=0, xmax=6.0,
xtick={0,...,5},
ymin=0,
ymax=11,
ytick distance=2,
%every axis x label/.style={at={(axis description cs:-0.3,-0.6)},anchor=west,},
%every axis y label/.style={at={(current axis.north west)},above=3mm},
xlabel near ticks,
ylabel near ticks,
%yticklabel style={/pgf/number format/.cd, fixed, fixed zerofill, precision=0, /tikz/.cd},
%scaled y ticks=base 10:0,
bar width = 5pt,
xlabel={\#Samples per iteration},
ylabel={Average \#iterations},
legend style={
	font=\small,
	at={(1.4,1.0)},
	anchor=north east,
}
]
\addplot +[ybar, fill=lightblue, draw=black, area legend] table[x={alpha},y={y1}]{bar.dat};
%\addplot +[ybar, area legend] table[x={alpha},y={y2}]{bar.dat};
%\addplot +[ybar, area legend] table[x={alpha},y={y3}]{bar.dat};
%\legend {\#iterations, \#adders, hypervolume error};
%\draw (axis cs:1,0) -- (axis cs:1,3) node [above] {Ours};
%\draw (axis cs:0,3) node [above] {Ours \&};
%\draw (axis cs:5,10) -- (axis cs:5,11);
%\draw (axis cs:5,11) -- (axis cs:5,12) node [above] {\cite{TPL_ISQED2013_Chen}};
\end{axis}
\end{tikzpicture}}  \hspace{.2in}
    \subfloat[]{\pgfplotsset{
	width =0.3\textwidth,
	height=0.23\textwidth
}

\begin{tikzpicture}[scale=1]
\begin{axis}[minor tick num=0,
ybar,
xmin=0, xmax=6.0,
xtick={0,...,5},
ymin=950,
ymax=1000,
ytick distance=10,
%yticklabel style={/pgf/number format/.cd, fixed, fixed zerofill, precision=0, /tikz/.cd},
%scaled y ticks=base 10:0,
bar width = 5pt,
xlabel={\#Samples per iteration},
ylabel={Average \#adders},
ylabel near ticks,
%every axis x label/.style={at={(axis description cs:-0.3,-0.6)},anchor=west,},
xlabel near ticks,
%every axis y label/.style={at={(current axis.north west)},above=3mm},
%legend entries={\#iterations, \#adders, HV error},
legend style={
	font=\small,
	at={(1.4,1.0)},
	anchor=north east,
}
]
%\addplot +[ybar, area legend] table[x={alpha},y={y1}]{bar.dat};
\addplot +[ybar, fill = lightblue, draw=black, area legend] table[x={alpha},y={y2}]{bar.dat};
%\addplot [ybar,draw=black, area legend] table[x={alpha},y={y2}]{bar.dat};
%\addplot +[ybar, area legend] table[x={alpha},y={y3}]{bar.dat};
%\legend {\#iterations, \#adders, hypervolume error};
%\draw (axis cs:1,0) -- (axis cs:1,3) node [above] {Ours};
%\draw (axis cs:0,3) node [above] {Ours \&};
%\draw (axis cs:5,10) -- (axis cs:5,11);
%\draw (axis cs:5,11) -- (axis cs:5,12) node [above] {\cite{TPL_ISQED2013_Chen}};
\end{axis}
\end{tikzpicture}}  \hspace{.2in}
    \subfloat[]{\pgfplotsset{
	width =0.3\textwidth,
	height=0.23\textwidth
}

\begin{tikzpicture}[scale=1]
\begin{axis}[minor tick num=0,
ybar,
xmin=0, xmax=6.0,
xtick={0,...,5},
ymin=0.00,
ymax=0.08,
ytick distance=0.02,
scaled ticks=false,
yticklabel style={/pgf/number format/.cd, fixed, fixed zerofill, precision=3, /tikz/.cd},
%scaled y ticks=base 10:0,
bar width = 5pt,
xlabel={\#Samples per iteration},
ylabel={Average hv error},
%every axis x label/.style={at={(axis description cs:-0.3,-0.6)},anchor=west,},
xlabel near ticks,
ylabel near ticks,
%every axis y label/.style={at={(current axis.north west)},above=3mm},
%legend entries={\#iterations, \#adders, HV error},
legend style={
	font=\small,
	at={(1.4,1.0)},
	anchor=north east,
}
]
%\addplot +[ybar, area legend] table[x={alpha},y={y1}]{bar.dat};
%\addplot +[ybar, area legend] table[x={alpha},y={y2}]{bar.dat};
\addplot +[ybar, fill=lightblue, draw=black, area legend] table[x={alpha},y={y3}]{bar.dat};
%\legend {\#iterations, \#adders, hypervolume error};
%\draw (axis cs:1,0) -- (axis cs:1,3) node [above] {Ours};
%\draw (axis cs:0,3) node [above] {Ours \&};
%\draw (axis cs:5,10) -- (axis cs:5,11);
%\draw (axis cs:5,11) -- (axis cs:5,12) node [above] {\cite{TPL_ISQED2013_Chen}};
\end{axis}
\end{tikzpicture}}
    \caption{Comparison among different number of samples per iteration.}
    \label{fig:result-sample}
\end{figure*}

\subsection{Comparison on Different Sampling Strategies in PAL}
\label{sec:diff-sample}
In the sampling stage of PAL, the number of instances to be sampled has impact on the runtime since the EDA flow is required to obtain the real value for area, power and delay.
The less instances we sample in each iteration, the more iterations are needed to ensure the PAL process converge, which is more likely to result in less samples in total.
The more instances we sample in each iteration, the less iterations are needed.
However, the total number of sampled instances would be large.
In practice, the runtime cost of running EDA flow can be reduced by parallel execution if there are multiple licenses available.
In this section, we explore the effect of different sampling strategies in terms of the total runtime and the quality of Pareto frontier in practical scenarios.

The results for different sampling strategies are listed in \Cref{fig:result-sample}.
Since the EDA flow for synthesis, placement and routing takes up the most significant part of the total runtime cost, the key factor is the number of adders which needs to be through EDA tool flow.
If we have multiple machines available for the EDA tool flow, the runtime is determined by the total number of iterations as long as the number of samples does not exceed the number of machines.
From the result, it can be seen that we can obtain the Pareto-frontier with comparable quality, using less runtime.

Note that when the sample size increases from 1 to 5, the average hypervolume error increases from 0.056 to about 0.070, which is still less than 0.154 (average hypervolume error achieved in $\alpha$-sweep approach).
Therefore, batch sampling can not only take care of parallel synthesizing but also achieve better quality for Pareto frontier than $\alpha$-sweep, which can also show the advantages of the PAL.

\iffalse
\begin{table}[tb!]
    \caption{Comparison among different sampling strategies} \label{tab:sample}
    \centering
    \resizebox{9cm}{!} {
        \begin{tabular}{|c|c|c|c|}
            \hline
            \multirow{2}{1.5cm}{\centering \#samples per iteration} & \multirow{2}{1cm}{\centering \#total iterations} & \multirow{2}{1cm}{\centering \#total samples} & \multirow{2}{1.5cm}{\centering Hypervolume error} \\
            %\hline
            &  &  &    \\   \hline
            1 & 5.7 & 390.7 & 0.080   \\   \hline
            2 & 4.2 & 383.9 & 0.082   \\   \hline
            3 & 4.0 & 382.9 & 0.083   \\   \hline
            4 & 3.6 & 381.4 & 0.084   \\   \hline
            5 & 3.5 & 378.4 & 0.086   \\  
            \hline
        \end{tabular}
    }
\end{table}
\fi

\subsection{Adder Performance Comparison}
\label{sec:was}
Finally, we compare our explored adders against DesignWare adders, legacy adders, such as Kogge-Stone, Sklansky, as well as a state-of-the-art adder synthesis algorithm in \Cref{tab:64}.
%The DesignWare adders are synthesized with the same setting options that are used in generating the solution space in \Cref{fig:wider}. 
Since our approach generates numerous solutions, it is not feasible to perform a one-to-one comparison.
Instead for each of the solution points in regular adders and \cite{ADDER_ASPDAC2015_Roy}, we have picked the Pareto points from our solution set which are able to excel them in all metrics.
For instance, $P_1$ could provide around 8$ps$ better delay with respectively $14\%$ and $12\%$ lesser area and power over Kogge-Stone adder.
The DesignWare adders are synthesized from behavioral description of adder (Y = A + B) with the $16$ configurations of tool settings (Combination of $4$ target delay and $4$ utilization values) that are used in generating the \Cref{fig:wider}. 
We pick the one with best delay, denoted by ``DesignWare'' in \Cref{tab:64}.
The same pareto point $P_1$ dominates that solution by providing around $7.5ps$ better delay, $14\%$ lesser area, and $15\%$ lesser energy.
%Since our approach generates numerous solutions, it is not feasible to perform a one-to-one comparison.
%Instead for each of the solution points in regular adders and \cite{ADDER_ASPDAC2015_Roy}, we have picked the Pareto points from our solution set which are able to excel them in all metrics.
%For instance, $P_1$ could provide around 8$ps$ better delay with respectively $14\%$ and $12\%$ lesser area and power over Kogge-Stone adder.
%We pick up the best delay solution for the comparison.
For \cite{ADDER_ASPDAC2015_Roy} we pick the best delay solution.
Note for a fixed $mfo$, \cite{ADDER_ASPDAC2015_Roy} can give prefix network with smaller size, but this approach only provides a limited set of prefix structures.
As a result, it is hard for \cite{ADDER_ASPDAC2015_Roy} to explore the full physical design space of adders by machine learning.
It should be stressed that \cite{ADDER_ASPDAC2015_Roy} beats the custom adders implemented in an industrial design,
and our methodology is able to excel the adders generated by the algorithm presented in \cite{ADDER_ASPDAC2015_Roy}.

\begin{table}[tb!]
    \caption{Comparison with other approaches for $64$ bit adders} \label{tab:64}
    \centering
    \resizebox{9cm}{!} {
        \begin{tabular}{|c|c|c|c|c|}
            \hline
            Method & Delay ($ps$) & Area (${\mu m}^2$) & Energy ($fJ$/op) \\
            \hline \hline
            DesignWare & 346.5 & 2531.3 & 8160 \\
            \textbf{{Ours ($P_1$)}}  & {339.0} & {2180.8}  &  {6930} \\
            \hline
            Kogge-Stone & 347.9 &  2563.7 & 8780 \\
            \textbf{{Ours ($P_1$)}}  & {339.0} & {2180.8}  &  {6930} \\
            \hline
            Sklansky & 356.1 & 1792.5 & 6100 \\
            \textbf{Ours ($P_2$)} & 353.0 & 1753.0 & 5900 \\
            \hline
            \cite{ADDER_ASPDAC2015_Roy} & 348.7 & 1971.4 & 6980 \\
            %\cite{ADDER_ASPDAC2015_Roy} & 440.7 & 1668.0 & 144.4 \\
            \textbf{{Ours ($P_3$)}} & {343.0} & {1912.6} & {6390} \\
            %\textbf{Our} & \textbf{436.6} & \textbf{} & \textbf{147.8} \\
            \hline
        \end{tabular}
    }
\end{table}

\iffalse
\section{Discussions} \label{sec:discuss}
In this work, we have focused on applying machine learning techniques to mapping from architectural solution space to physical solution space for high performance adders.
In doing so, the features from (i) architectural stage, (ii) synthesis, and (iii) physical design tool settings have been considered.
Although this work targets and is limited to only adders, similar machine learning based methodology can be adopted in exploring other complex arithmetic circuits, such as multipliers etc.
However, it may be difficult to always find the regularity in larger general circuits. 
This may cause a hindrance to find appropriate features, but in that case, deep neural network may be explored for automatic feature learning.
\fi

\section{Conclusion} \label{sec:conclu}
\iffalse
This paper presents a novel methodology of machine learning guided design space exploration for high-performance prefix adders. 
We have successfully demonstrated that our approach, to search for optimal delay vs.~power/area Pareto frontier, 
could potentially achieve around 40$\times$ speed-up in turn-around-time (TAT) over exhaustive design space exploration. 
To the best of our knowledge, this is the first work to bridge the gap between architectural and physical solution space for parallel prefix adders. 
With increasing design complexity and cross-layer uncertainties, we anticipate that our methodology will become more and more relevant, 
and it will be adopted for other designs in recent future.   
\fi

This paper presents a novel methodology of machine learning guided design space exploration for power efficient high-performance prefix adders. 
We have successfully demonstrated the effectiveness of our learning models, developed by training with quasi-random sampled data and features encapsulating architectural and tool attributes.
In addition, an active learning approach is applied to ease the demand of labeled data and achieves even better Pareto frontier.
Our adder synthesis algorithm is able to generate a wider solution space in comparison to a state-of-the-art algorithm, 
and when integrated with the learning model, could provide a remarkable performance vs.~power vs.~area Pareto frontier over a large representative solution space.
%Although we have currently focused on $64$ bit adders with logic level $6$ (to target high-performance adders), 
%our methodology is general to explore adders of other bit-widths and logic-levels.   
To the best of our knowledge, this is the first work to bridge the gap between architectural and physical solution space for parallel prefix adders.
%In microprocessors, binary adders are often the building blocks, and the critical path delay of adder block may dictate the overall performance of the microprocessors.
%So the power gains in adder design could potentially lead to better system energy efficiency.

%In future we plan to extend our methodology to address the gap between architectural/logic stage and physical design for other larger instance circuits.
%However, it may be difficult to always find the regularity in the circuit structure like prefix adders.
%This can cause a hindrance to find the features, but in that case, multilayer neural network can be explored for efficient feature learning.
%In addition, physical synthesis options may also impact the performance of the adders.
%Exploring the physical synthesis options may help to make the proposed methodology more general.

%\balance
{
\bibliographystyle{IEEEtran}
\bibliography{./ref/Top-sim,./ref/Software,./ref/LEARN,./ref/Synthesis,./ref/PD,./ref/Timing,./ref/HSD,./ref/CAD}
}

\end{document}